\documentclass[pdflatex,sn-basic,Numbered]{sn-jnl}

\usepackage{graphicx}
\usepackage{amsmath,amssymb,amsfonts}
\usepackage{amsthm}
\usepackage{tikz}
\usetikzlibrary{positioning, arrows.meta, fit, calc, shapes.geometric, backgrounds}
\usepackage[title]{appendix}
\usepackage{xcolor}
\usepackage{booktabs}
\usepackage{algorithm}
\usepackage{algorithmicx}
\usepackage{algpseudocode}
\usepackage{listings}

\lstdefinelanguage{MLIR}{
  morekeywords={func, return, arith, scf, memref, llvm, attributes, qubit,
    alloc, h, cnot, measure},
  sensitive=true,
  morecomment=[l]{//},
  morestring=[b]"
}
\let\orcidlogo\relax
\usepackage{orcidlink}

\begin{document}

\title{QWeaver: Quantum Circuit Cut Workflow with MLIR-based Compiler, Runtime, and Weaver Hybrid Tasks}

\author*{\fnm{Narasinga Rao} \sur{Miniskar}\,\scalebox{0.5}{\orcidlink{0000-0001-8259-8891}}}%
\email{miniskarnr@ornl.gov}
\equalcont{These authors contributed equally to this work.}

\author{\fnm{Elaine} \sur{Wong}\,\scalebox{0.5}{\orcidlink{0000-0003-1354-2020}}}
\email{wongey@ornl.gov}
\equalcont{These authors contributed equally to this work.}

\author{\fnm{Vicente} \sur{Leyton-Ortega}\,\scalebox{0.5}{\orcidlink{0000-0003-4255-9892}}}
\email{leytonorteva@ornl.gov}
\equalcont{These authors contributed equally to this work.}

\author{\fnm{Sharmin} \sur{Afrose}\,\scalebox{0.5}{\orcidlink{0000-0003-1835-773X}}}
\email{afroses@ornl.gov}
\equalcont{These authors contributed equally to this work.}

\author{\fnm{Jeffrey S.} \sur{Vetter}\,\scalebox{0.5}{\orcidlink{0000-0002-2449-6720}}}
\email{vetter@ornl.gov}

\author{\fnm{Travis S.} \sur{Humble}\,\scalebox{0.5}{\orcidlink{0000-0002-9449-0498}}}
\email{humblets@ornl.gov}

\affil{
  \orgname{Oak Ridge National Laboratory},
  \orgaddress{
    \city{Oak Ridge},
    \postcode{37830},
    \state{Tennessee},
    \country{USA}
  }
}

\begingroup
\renewcommand{\thefootnote}{}
\footnotetext{\textbf{Notice:} This manuscript has been authored by UT-Battelle, LLC, under contract DE-AC05-00OR22725 with the US Department of Energy (DOE). The US government retains and the publisher, by accepting the article for publication, acknowledges that the US government retains a nonexclusive, paid-up, irrevocable, worldwide license to publish or reproduce the published form of this manuscript, or allow others to do so, for US government purposes. DOE will provide public access to these results of federally sponsored research in accordance with the DOE Public Access Plan ( https://www.energy.gov/doe-public-access-plan ). 
}
\addtocounter{footnote}{-1}
\endgroup
\abstract{
Quantum computers available today are constrained by limited qubit counts and high noise levels, making the direct execution of large quantum circuits infeasible on a single device, and while quantum circuit cutting addresses this by partitioning a large circuit into smaller subcircuits for independent execution, orchestrating the full end-to-end workflow remains a significant programming challenge. In this paper, we present \textit{QWeaver}, a task-graph-based programming and orchestration workflow framework for quantum circuit cutting that accepts OpenQASM quantum circuits as input and constructs a directed acyclic task graph. In this graph, the Multi-Level Intermediate Representation (MLIR)-based circuit cut is the first task, responsible for analyzing and partitioning the input circuit into executable fragment representations; the generated fragments are then handed to a set of quantum circuit execution tasks that depend on the cut task and can be dispatched concurrently across available quantum backends via the Quantum Intelligent Runtime System (Q-IRIS). Finally, a weaving task, dependent on all execution tasks, performs classical post-processing that applies quasi-probability reconstruction techniques to combine the sampled measurement outcomes of all fragments and recover an accurate approximation of the original circuit's output distribution.We describe the architecture of QWeaver, its integration with Q-IRIS and the execution engine for the Quantum Intermediate Representation (QIR-EE), validate end-to-end reconstruction on GHZ and Harris--Ross instances, and evaluate systems and width trade-offs on structured GHZ simulator workloads.
}


\maketitle

\section{Introduction}
\label{sec:introduction}

Practical quantum advantage over classical computers remains out of reach for most applications, not because quantum algorithms lack promise, but because the hardware available today falls well short of what those algorithms require. Current quantum processors, often described as Noisy Intermediate-Scale Quantum (NISQ) devices~\cite{Preskill2018}, offer tens to a few hundred qubits, and even circuits that nominally fit a device are constrained by its connectivity and by the depth decoherence permits. In practice, the circuits of near term interest are frequently somewhat, not astronomically, wider than the devices or simulators available to run them; an application circuit of forty to eighty qubits against a device of twenty or thirty, or a simulation workload just past the memory wall that limits single-node state-vector simulation to roughly 35-40 qubits. Circuits like these are out of reach as a whole, yet not by order of magnitude. 

Quantum circuit cutting is one practical response to this mismatch. The idea, first put on rigorous footing by ~\cite{Peng2020}, is to cut a large quantum circuit across a small number of wires, replacing those cuts with a set of local operations and measurements, then combine the results from the resulting smaller subcircuits, {\it the fragments}, using classical post-processing to reconstruct the probability distribution of the original circuit's output. The overhead is exponential in the number of cuts, so the technique is most useful when a circuit has a sparse entanglement structure that allows a small number of cuts to break it into independently executable fragments. This overhead defines the practical limits of circuit cutting: it is not a route to fault-tolerant-scale algorithms, whose qubit requirements cannot be bridged by sampling-based decomposition, but rather a constant-factor extension of device or simulator width, achieved through additional classical computation for circuits whose structure requires relatively few cuts.
A second consequence, central to this paper, is that the fragment programs are mutually independent. One wide circuit therefore becomes a large set of concurrent, schedulable narrow tasks, a shape that matches multi-QPU and high performance computing (HPC) resources far better than the original monolithic circuit.  Several groups have since explored how to make this overhead more tractable~\cite{Mitarai2021, Lowe2023} and how to build tooling around it~\cite{Tang2021, Piveteau2024}.

The tooling gap, however, is real. Cutting a circuit is only one step in a longer pipeline: (1) the input circuit must be compiled and analyzed to identify suitable cut points; (2) each resulting fragment must be lowered into an executable form; (3) the fragments must be dispatched to one or more quantum backends; and (4) the resulting measurement samples must be processed by a reconstruction procedure to produce the final output distribution. Running this pipeline end-to-end in a way that is both correct and efficient, especially when multiple QPUs are available, requires a programming model that captures the dependencies between these stages and schedules them accordingly. Existing tools address individual stages of this pipeline, such as cut placement, fragment generation, or reconstruction. To our knowledge, none expresses all three as scheduled tasks with explicit data dependencies. We do not claim a performance advantage over these tools, and a controlled quantitative comparison is left to future work.\label{par:workflow-scope}

MLIR~\cite{Lattner2021} (Multi-Level Intermediate Representation) has emerged as a natural foundation for quantum compiler infrastructure because its dialect system lets researchers compose transformations across levels of abstraction, from high-level quantum semantics down to hardware-specific representations without abandoning a single coherent IR. Several quantum computing projects have adopted MLIR dialects~\cite{Ittah2024Catalyst,qecompiler,cudaq} to represent and transform circuits expressed in OpenQASM~3~\cite{Cross2022}, a widely used quantum assembly language that supports rich classical control and is increasingly treated as the lingua franca for near-term algorithm expression. The Quantum Intermediate Representation (QIR)~\cite{QIR2021} provides a complementary LLVM-based format for lowering these circuits to a form suitable for backend execution.

In this paper we present QWeaver, a task-graph-based programming and orchestration framework that ties together the full circuit-cutting pipeline, an MLIR-based cut placement and compilation, with fragment execution on quantum backends, and classical result reconstruction, as a single coherent workflow. Where our earlier Q-IRIS integration executed externally prepared circuits on a single QPU~\cite{Miniskar2026b}, QWeaver makes cutting itself a compiler task inside the scheduled graph, and its execution tasks can be dispatched concurrently across whatever quantum backends, physical or simulated, are available. QWeaver takes an OpenQASM~3 circuit as input and builds a directed acyclic task graph over three stages. The first task performs the MLIR-based circuit cut, analyzing the input circuit and emitting a set of executable fragment programs together with an execution manifest. A set of quantum circuit execution tasks, each dependent on the cut task, then submits these programs to quantum backends through the QIR-EE execution engine~\cite{wong2025qiree} and collects the resulting measurement samples. Finally, a weaving task, dependent on all execution tasks, combines those samples using quasi-probability reconstruction~\cite{Peng2020, Mitarai2021} to recover an estimate of the original circuit's observables. The task graph is scheduled and executed by the Q-IRIS heterogeneous runtime~\cite{miniskar2026q, Miniskar2026b}, which handles dependency tracking, data movement between tasks, and concurrent dispatch across available devices. The contributions of this paper are not a new cutting decomposition, QPD method, intermediate representation, or heterogeneous scheduler, but their end-to-end integration in a data-dependent task graph in which quantum circuit cutting is a compiler task whose outputs directly determine the fragment-execution and reconstruction tasks. In more detail, it contributes: 

\begin{enumerate} 
\item A task-graph programming model for end-to-end quantum circuit cutting workflows, in which the cut, execute, and weave stages, the first a compiler pass and the last a classical reduction, appear as first-class tasks with explicit data dependencies. 
\item An MLIR-based circuit-cut pass, integrated with the Q-IRIS runtime and the QIR-EE execution engine, that analyzes and partitions the input circuit and lowers each fragment to standard QIR, enabling concurrent execution of fragment programs across multiple quantum backends, physical or simulated. 
\item A weaving layer that applies quasi-probability reconstruction to the collected fragment samples and produces an estimate of the original circuit's observables, with the full output distribution available as a second, signed target type. \item An experimental evaluation on GHZ circuits and a weighted MaxCut instance derived from the Harris--Ross railway network, spanning six simulator configurations and an IQM hardware endpoint, that validates reconstructed observables against exact baselines and characterizes end-to-end accuracy, runtime, and feasibility as circuit width, cut count, and per-fragment backend-qubit capacity vary. 
\end{enumerate}

The remainder of the paper is organized as follows. Section~\ref{sec:background} provides background on MLIR, the Q-IRIS runtime, and quantum circuit cutting.  Section~\ref{sec:qweaver} describes the architecture of QWeaver, from the MLIR-based circuit cut pass through the QIR-EE execution layer to the weaving stage. Section~\ref{sec:cost-model} develops a first-order cost model to help understand when circuit cutting will prove useful. Section~\ref{sec:results} presents experimental results, Section~\ref{sec:related-work} discusses related work, and Section~\ref{sec:conclusion} concludes.

\section{Background}
\label{sec:background}

\subsection{MLIR}

MLIR~\cite{Lattner2021} makes the IR itself extensible: a small core of operations, regions, values, types, and attributes provides the substrate, and \emph{dialects} on top of that substrate capture domain-specific semantics. Built-in dialects span the abstraction stack from \texttt{arith} and \texttt{scf} through \texttt{affine} to \texttt{llvm}, and they coexist in a single module so an MLIR program descends through a sequence of mixed-dialect representations under \emph{progressive lowering}. Transformations are written as \emph{passes} whose declarative pattern-rewrite rules compose into dialect conversions, and cross-dialect interfaces (memory effects, side-effect freedom, type inference) let generic passes reason about user-defined operations without hard-coding each dialect. Listing~\ref{lst:mlir-bell} shows what a value-semantic quantum dialect looks like in this style: each gate consumes its input qubit static single-assignment (SSA) values and produces fresh ones, so quantum dataflow is explicit in the IR and the no-cloning property is enforced by SSA itself.

\begin{lstlisting}[float,label=lst:mlir-bell,caption={Bell-pair preparation in a representative value-semantic quantum MLIR dialect. Each gate consumes and produces qubit static single-assignment (SSA) values, making quantum dataflow explicit in the IR.}]
func.func @bell() {
  %q0   = quantum.alloc : !quantum.qubit
  %q1   = quantum.alloc : !quantum.qubit
  %q0_h = quantum.h    %q0 : !quantum.qubit
  %q0_c, %q1_c = quantum.cnot %q0_h, %q1
      : !quantum.qubit, !quantum.qubit
  %m0 = quantum.measure %q0_c : i1
  %m1 = quantum.measure %q1_c : i1
  return
}
\end{lstlisting}

This combination fits quantum compilation well, where a program is naturally expressed at the level of gates and qubit registers but must still be lowered through gate-set decomposition, qubit routing under connectivity constraints, pulse scheduling, and backend code emission. Several MLIR-based quantum compilers already exploit this property: McCaskey and Nguyen~\cite{McCaskey2021} define a quantum dialect that ingests OpenQASM and lowers to LLVM~IR following the QIR~\cite{QIR2021} specification; QSSA~\cite{Peduri2022} represents qubits as SSA values, statically enforcing no-cloning and enabling classical dataflow analyses on quantum programs; QIRO~\cite{Ittah2022} pairs memory-semantic and value-semantic dialects to expose quantum dataflow to standard MLIR passes; and Catalyst~\cite{Ittah2024Catalyst} extends MLIR to just-in-time compile PennyLane's hybrid quantum-classical Python programs. Circuit cutting exists in this ecosystem, but above the compiler: PennyLane places cuts on the quantum circuit's graph with KaHyPar and marks them with placeholder measure and prepare nodes~\cite{pennylanecut}, and the fragments return to the compiler only as new circuit objects. Section~\ref{subsec:cut-task} instead describes a circuit-cut pass over a quantum-circuit dialect that consumes OpenQASM~3, analyzes entanglement structure to identify cut points, and emits fragment representations that downstream stages execute and recombine.

\subsection{Q-IRIS: IRIS with Quantum support}
\label{subsec:qiris}

\begin{figure*}[t]
  \centering
  \includegraphics[width=\textwidth]{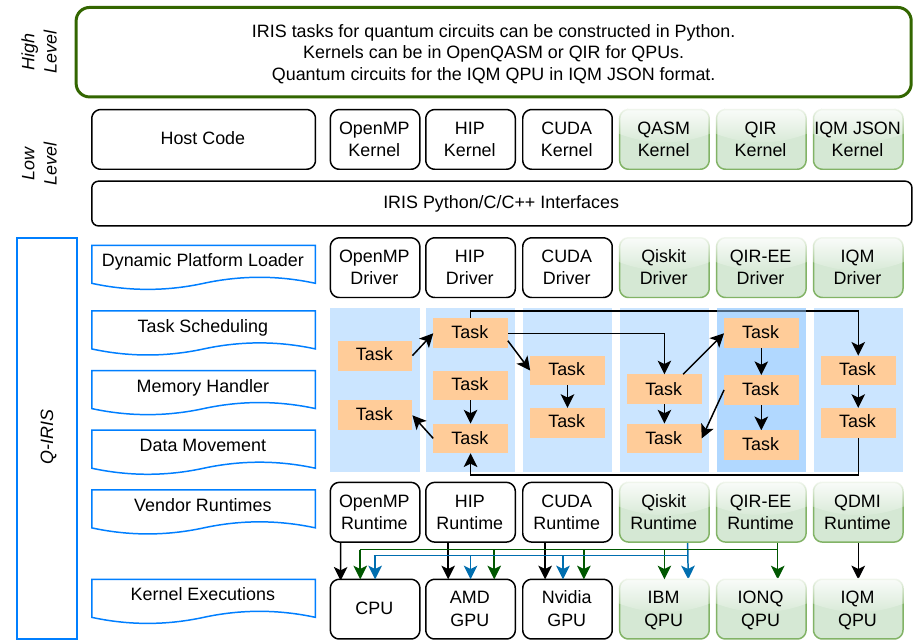}
  \caption{IRIS, the Intelligent Runtime System, with support for quantum
    processing units. The tightly coupled integration of QIR-EE and the QDMI into the IRIS runtime is highlighted in the green boxes.}
  \label{fig:iris-background}
\end{figure*}

QWeaver's runtime layer, Q-IRIS, extends the IRIS heterogeneous runtime with a quantum execution path; this subsection reviews IRIS itself, and Section~\ref{sec:qiree} introduces the execution engine that provides that path. IRIS~\cite{kim2024iris,miniskar2024iris} is a task-based runtime that brings a broad spectrum of compute devices (CPUs, NVIDIA and AMD GPUs, Xilinx and Intel FPGAs, and DSPs) under one concise, task-based API. Developers create and submit work items whose kernels are written in the model appropriate to each target (CUDA, HIP, OpenCL, HLS C++), and the runtime supplies multi-stream asynchronous execution~\cite{miniskar2025iris}, intelligent data orchestration~\cite{miniskar2023iris} that automatically migrates data between host and accelerator memories, configurable scheduling policies that dynamically decide where each task runs~\cite{johnston2024iris}, and dynamic compilation that builds and optimizes kernels lazily at execution time. Its plug-in device model already backs high-level frameworks such as OpenACC, SYCL, Python, and Julia, and, most relevant here, was recently extended to treat quantum processors as devices that execute programs expressed in the Quantum Intermediate Representation (QIR)~\cite{miniskar2026q}, as illustrated in Fig.~\ref{fig:iris-background}. That early integration was limited to a single QPU because the underlying execution engine (QIR-EE, reviewed next) was not thread-safe; QWeaver builds on the full multi-QPU integration presented in Section~\ref{sec:qweaver}. We have extended Q-IRIS with QDMI (Quantum Device Management Interface) \cite{qdmi} to support QPUs such as IQM Pathfinder located at Oak Ridge National Laboratory. 

\subsection{QIR-EE}
\label{sec:qiree}
The QIR Execution Engine~\cite{wong2025qiree,qireerepo}
is a C++ library from ORNL that translates quantum programs written in the
Quantum Intermediate Representation (QIR)~\cite{qirspec,QIR2021} into
executable code for both simulators and real hardware. Given a QIR
\texttt{.ll} file, a chosen accelerator target, and a shot count, QIR-EE uses
LLVM's runtime infrastructure to process the program's classical control
flow while exposing hooks that bind backend-specific quantum operations. It
can reach Google's QSIM state-vector simulator~\cite{qsim} directly, and it
routes simulator and hardware access through the XACC
framework~\cite{xaccrepo}. On its own, however, QIR-EE cannot launch multiple
QIR programs concurrently across accelerators, a limitation that the IRIS
integration removes.

Q-IRIS is the union of these two systems: IRIS extended with a QIR-EE device
driver so that QIR programs become first-class tasks that the scheduler can
dispatch, concurrently and across multiple QPUs, alongside classical work.
The architecture of this integration, and the driver mechanism that makes it
thread-safe and multi-QPU capable, are the subject of
Section~\ref{sec:qweaver}.

\subsection{Wire cutting by quasi-probability decomposition}

Consider a qubit wire that carries a state from an upstream portion of a circuit to a downstream portion.  In the uncut circuit this connection is the single-qubit identity channel, it transmits an arbitrary state, including its correlations with the rest of the circuit. A wire cut removes that quantum connection and replaces it with operations that the two sides can perform separately.  The upstream side measures the boundary qubit, the downstream side prepares a prescribed state, and classical post-processing combines the resulting experiments so that their signed sum reproduces the identity channel~\cite{Peng2020, Mitarai2021}.

For a state $\rho$ arriving at one cut, this quasi-probability decomposition (QPD) has the form 

\begin{equation}\label{eq:identity}
    {\cal I} (\rho) = \sum_{\alpha \in {\cal A}} c_\alpha {\rm Tr} (M_\alpha \rho) \sigma_\alpha ,
\end{equation}

where $M_\alpha$ is the observable measured immediately before the cut, $\sigma_\alpha$ is the state prepared immediately after it, and $c_\alpha$ is a real coefficient. The equality is a linear identity, not a probabilistic mixture, meaning, some coefficients are negative, which is why the construction is called a {\it quasi-probability} decomposition. Operationally, ${\rm Tr}(M_\alpha \rho)$ is estimated from measurement outcomes, while $\sigma_\alpha$ is prepared in a separate circuit execution, so, no quantum state must pass between the two fragments.  

QWeaver uses the complete eight-term Pauli decomposition of a single wire cut \cite{Peng2020}. Its set {$\cal A$} contains two-identity-observable terms that prepare $|0\rangle$ and $|1\rangle$, each with coefficient $1/2$. For every $P \in \lbrace X,Y,Z \rbrace$ it also contains two terms, measure $P$ and prepare its $s \in \lbrace +1, -1 \rbrace$ eigenstate $|P_s\rangle$, with coefficient $s/2$. These eight signed measured-and-prepare contributions reconstruct an arbitrary single-qubit state through \eqref{eq:identity}. Appendix~\ref{app:pauli8} lists the eight terms explicitly and derives the corresponding reconstruction, experiment-count, and sampling overhead. 

With $k$ independent decomposed cuts, a boundary assignment $\alpha = (\alpha_1, \ldots, \alpha_k) \in \mathcal{A}^{\times k}$ selects one term at each cut and has coefficient $c_\alpha = \prod_{i=1}^k c_{\alpha_i}$. (See Figure~\ref{fig:app-pauli8}.) 
Removing the cut connections partitions the original circuit into $r$ {\it fragments}, structural subcircuits whose cut boundaries still denote abstract measurement and preparation operations. A {\it fragment program} is instead one concrete, executable realization of fragment $f$ under assignment $\alpha$, after those boundary operations have been selected and the circuit has been lowered to QIR.  The complete Pauli-8 plan therefore contains $8^k$ boundary assignments and $r 8^k$ fragment programs before pruning.

Executing the fragment programs produces local expectation values or output distributions. Classical reconstruction multiplies the compatible fragment contributions for each $\alpha$, applies $c_\alpha$, and sums over all assignments. \eqref{eq:identity} makes this reconstruction exact when the local quantities are known exactly, so finite-shot estimates add statistical error and can yield small negative reconstructed probabilities.  Wire cutting thus exchanges quantum width for classical and sampling work: the fragments may fit devices that cannot hold the original circuit, but the number of experiments and the estimator variance grow rapidly with the cut count (see Appendix~\ref{app:pauli8}).  This trade-off motivates both the minimal-cut compiler objective and the cut-execute-weave task graph developed in Section \ref{sec:qweaver}.

\section{QWeaver framework}
\label{sec:qweaver}

QWeaver turns the multi-stage circuit-cutting pipeline into a single \emph{directed acyclic task
graph} (DAG) that the Q-IRIS heterogeneous runtime schedules end to end.
The design follows one principle: every stage that the pipeline performs
once (analyze and cut the circuit, execute a fragment, recombine/weaving the
samples) is expressed as a first-class \emph{task} with explicit data
dependencies, and no stage manages device placement or concurrency itself.
Placement, dependency tracking, data movement, and concurrent dispatch are
delegated to Q-IRIS, exactly as they are for a classical heterogeneous
workload~\cite{kim2024iris,miniskar2024iris}. What makes the workflow hybrid
is only that some of the tasks in the graph are quantum circuit execution: they carry a QIR program to a QPU instead of a CPU/GPU kernel.

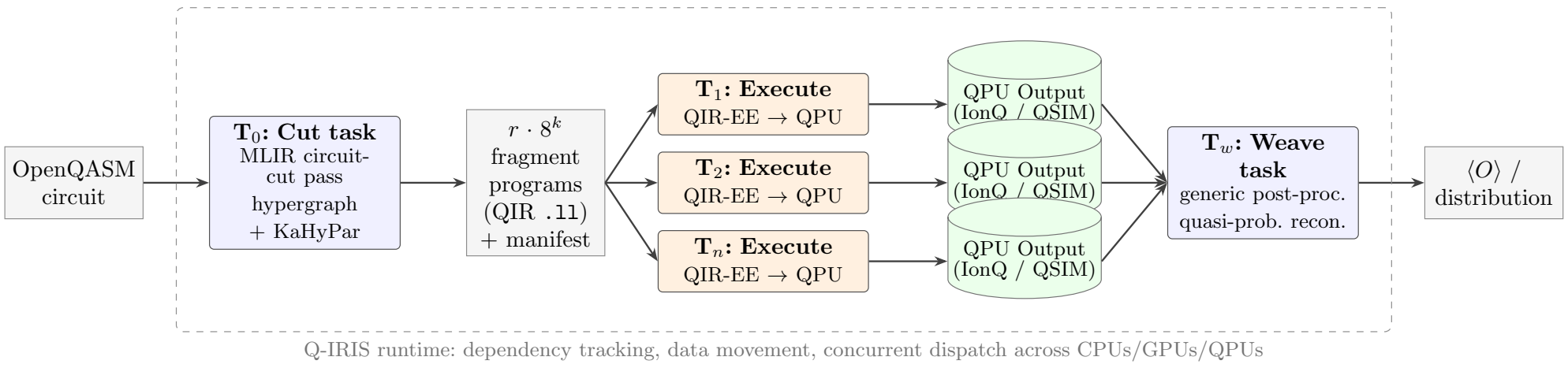
\begin{figure*}[t]
  \centering
  \resizebox{\textwidth}{!}{%
  \begin{tikzpicture}[
      node distance = 6mm and 10mm,
      font = \normalsize,
      task/.style  = {rectangle, rounded corners=2pt, draw=black!70,
                      fill=blue!6, align=center, minimum height=13mm,
                      text width=27mm, inner sep=3pt},
      qtask/.style = {rectangle, rounded corners=2pt, draw=black!70,
                      fill=orange!12, align=center, minimum height=9mm,
                      text width=30mm, inner sep=3pt},
      data/.style  = {rectangle, draw=black!45, fill=gray!8, align=center,
                      minimum height=11mm, text width=19mm, inner sep=3pt},
      dep/.style   = {-{Stealth[length=2mm]}, thick, draw=black!75},
      qpu/.style   = {cylinder, shape border rotate=90, aspect=0.25,
                      draw=black!55, fill=green!8, align=center,
                      minimum height=8mm, text width=22mm, inner sep=2pt,
                      font=\scriptsize},
    ]

    \node[data]  (qasm) {OpenQASM\\ circuit};

    \node[task]  (cut) [right=of qasm]
      {\textbf{T$_0$: Cut task}\\[1pt]
       \scriptsize MLIR circuit-cut pass\\
       \scriptsize hypergraph $+$ KaHyPar};

    \node[data]  (frag) [right=of cut]
      {$r \cdot 8^{k}$ fragment\\ programs (QIR \texttt{.ll})\\ $+$ manifest};

    \node[qtask] (e1) [right=8mm of frag, yshift=12mm]
      {\textbf{T$_1$: Execute}\\ \scriptsize QIR-EE $\rightarrow$ QPU};
    \node[qtask] (e2) [right=8mm of frag]
      {\textbf{T$_2$: Execute}\\ \scriptsize QIR-EE $\rightarrow$ QPU};
    \node[qtask] (en) [right=8mm of frag, yshift=-12mm]
      {\textbf{T$_n$: Execute}\\ \scriptsize QIR-EE $\rightarrow$ QPU};

    \node[qpu] (q1) [right=12mm of e1] {QPU Output\\ (IonQ / QSIM)};
    \node[qpu] (q2) [right=12mm of e2] {QPU Output\\ (IonQ / QSIM)};
    \node[qpu] (qn) [right=12mm of en] {QPU Output\\ (IonQ / QSIM)};

    \node[task]  (weave) [right=10mm of q2]
      {\textbf{T$_w$: Weave task}\\[1pt]
       \scriptsize generic post-proc.\\
       \scriptsize quasi-prob. recon.};

    \node[data]  (out) [right=of weave]
      {$\langle O \rangle$ /\\ distribution};

    \draw[dep] (qasm) -- (cut);
    \draw[dep] (cut)  -- (frag);
    \draw[dep] (frag.east) -- (e1.west);
    \draw[dep] (frag.east) -- (e2.west);
    \draw[dep] (frag.east) -- (en.west);
    \draw[dep] (e1) -- (q1);
    \draw[dep] (e2) -- (q2);
    \draw[dep] (en) -- (qn);
    \draw[dep] (q1.east) -- (weave.west);
    \draw[dep] (q2.east) -- (weave.west);
    \draw[dep] (qn.east) -- (weave.west);
    \draw[dep] (weave) -- (out);

    \node[draw=black!45, dashed, rounded corners=3pt, inner sep=5mm,
          fit=(cut)(frag)(e1)(en)(q1)(qn)(weave),
          label={[black!55,font=\scriptsize]below:%
            Q-IRIS runtime: dependency tracking, data movement, concurrent
            dispatch across CPUs/GPUs/QPUs}] {};

  \end{tikzpicture}%
  }
  \caption{End-to-end QWeaver workflow as a Q-IRIS task graph. The cut task
    T$_0$ compiles the OpenQASM~3 circuit through the MLIR circuit-cut pass
    and, for $k$ cuts and $r$ fragments, emits $r \cdot 8^{k}$ QIR fragment programs plus an execution manifest. The
    execute tasks T$_1\!\dots$T$_n$ fan out from T$_0$ and run the fragment
    programs concurrently on quantum backends (Google QSIM directly, or
    XACC-routed simulators and real QPUs such as IonQ) through the QIR-EE
    driver. The weave task T$_w$ depends on all execute tasks and runs a
    generic classical post-processing kernel that reconstructs the original
    observable or distribution. Q-IRIS owns all scheduling, placement, and
    data movement; QWeaver only declares the tasks and their dependencies.}
  \label{fig:qweaver-workflow}
\end{figure*}

Concretely, QWeaver ingests an OpenQASM~3 circuit and builds
the three-stage DAG shown in Fig.~\ref{fig:qweaver-workflow}~\cite{Cross2022}, which spans the
compilation and runtime phases. A single
\emph{cut task} lowers the circuit into an MLIR quantum-dialect module, runs
the MLIR-based circuit-cut compiler pass, and emits the resulting QIR fragment programs together with an execution manifest.
A fan-out of \emph{execute tasks}, each depending on the cut task, dispatches
those QIR programs to QPUs through the QIR-EE execution
engine~\cite{wong2025qiree,qireerepo} under Q-IRIS control and collects the
resulting measurement samples. Additionally, Q-IRIS can also execute quantum circuits written in IQM JSON files via QDMI. A final \emph{weave task}, depending on all
execute tasks, runs a generic classical post-processing kernel that applies
quasi-probability reconstruction~\cite{Mitarai2021,harada2024doubly} to
stitch the fragment samples back into an estimate of the original circuit's
observable or output distribution. The remainder of this section describes
the task-graph programming model (Section~\ref{subsec:taskgraph}) and then
each of the three task kinds in turn (Sections~\ref{subsec:cut-task}--%
\ref{subsec:weave-task}).

\subsection{Task-graph programming model}
\label{subsec:taskgraph}
Throughout this section, $k$ denotes the number of cut interfaces and $r$ denotes the number of resulting fragments.  
QWeaver expresses the workflow as a Q-IRIS task graph rather than an
imperative script for two reasons that are specific to circuit cutting.
First, the fan-out is data-dependent: $k$ cut interfaces produce $8^k$ Pauli-8 reconstruction rows, each requiring one execution of each of the $r$ fragments. A task graph lets the cut task emit its
manifest and let the runtime spawn the corresponding $r8^k$ execute tasks instead
of hard-coding a fan-out width. Second, the stages have sharply different
resource affinities: the cut task is a classical compiler pass, the execute
tasks are quantum, and the weave task is a classical reduction over the
collected measurement data, so binding them to devices manually would be
brittle. By declaring
each stage as a task with typed inputs and outputs, QWeaver hands Q-IRIS the
freedom to place the compiler and post-processing tasks on CPU cores, to
place the execute tasks on whichever QPUs (real or simulated) are available,
and to overlap them under its asynchronous, multi-stream
scheduler~\cite{miniskar2025iris,johnston2024iris}.

Data flows between tasks through the Q-IRIS heterogeneous-memory
model~\cite{miniskar2023iris}, which migrates buffers between host and device
memory automatically and maps QPU outputs directly to NumPy objects on the
host. This is what lets the weave task consume the fragment histograms as
ordinary arrays without any explicit copy management: the execute tasks
write expectation values and bitstring counts into Q-IRIS memory, and the
runtime materializes them where the classical post-processing kernel runs.
The three task kinds, described next, are therefore the only application
code QWeaver contributes; everything about \emph{when} and \emph{where} they
run is the runtime's responsibility.

\subsection{The cut task: MLIR circuit cutting}
\label{subsec:cut-task}

The cut task is the single entry point of the workflow and the only task that invokes the compiler. It raises the input circuit into the MLIR quantum dialect, partitions it, and emits one QIR fragment program per fragment per cut experiment ($r\,8^{k}$ programs) together with an execution manifest. Algorithm~\ref{alg:qcut} states the full workflow; this subsection describes its cut phase, and Sections~\ref{subsec:execute-task} and~\ref{subsec:weave-task} the other two. The objective is set by the stages that follow: the number of cut wires fixes how many fragment programs this task emits, so minimizing it minimizes both the fan-out the runtime must schedule and the sum the weave task must evaluate.

\paragraph{Canonical wires} A circuit is worth cutting only where it entangles qubits, since single-qubit rotations create no cross-qubit dependence and cannot force a wire to be cut. The pass therefore walks the module once in program order and reduces it to its two-qubit interactions. Two properties of the SSA form complicate that walk. A logical qubit is renamed at every gate it passes through, because each gate consumes a \texttt{!quantum.bit} value and produces a fresh one, and the same physical wire may be threaded through wire-preserving operations such as \textsc{swap}. The pass traces each qubit value backward through single-qubit pass-throughs and wire-preserving permutations until it reaches an origin, a block argument or an \texttt{extract}, and assigns that origin a canonical integer index. Two gate operands carry the same index exactly when they act on the same physical qubit, however the IR has renamed them.

\paragraph{Hypergraph and objective} With canonical indices in hand the pass
builds a hypergraph $H = (V, E)$ from the entangling structure alone. Each
two-qubit gate acting on two distinct canonical qubits becomes a vertex of
unit weight, and each qubit that such a gate touches becomes a hyperedge whose
pins are the gate vertices incident on it. A qubit wire is therefore
represented by the set of interactions strung along it, which is exactly the
object a wire cut severs. KaHyPar~\cite{kahypar} partitions $V$ into $r$
blocks under the connectivity objective \texttt{km1},
\begin{equation}
  \label{eq:km1}
  \mathrm{km1}(\Pi) \;=\; \sum_{e \in E} \big(\lambda(e) - 1\big)\, \omega(e) ,
\end{equation}
where $\lambda(e)$ is the number of blocks hyperedge $e$ touches and
$\omega(e)$ its weight, uniform in our construction. For a qubit hyperedge the
term $\lambda(e) - 1$ counts the partition boundaries that wire must cross, so
Eq.~\eqref{eq:km1} is a direct proxy for the cut count. Every block is held
within an imbalance tolerance $\varepsilon = 0.03$ of $\lceil |V| / r \rceil$,
which stops the partitioner from driving the cut to zero by dumping every gate
into one fragment, and a fixed seed makes the partition reproducible.

All experiments in this work use uniform hyperedge weights, so the current objective minimizes the structural wire-cut count under the fragment-balance constraint. It is not noise- or entanglement-aware: partitions with the same number of cuts may have different fragment depths, estimator variances, entanglement across their boundaries, and backend error rates, and therefore different reconstruction difficulty. The hypergraph formulation can incorporate such information through nonuniform hyperedge weights, but evaluating those weighting models is left to future work.

\paragraph{Cuts and fragments} The partition labels gates, not wires, so two
steps recover a cuttable circuit. For each qubit the pass lists its incident
gates in program order and inspects consecutive pairs: a pair whose gates land
in different blocks marks a wire cut, since that wire must be measured after
the first gate in its source fragment and re-prepared before the second in its
destination. Qubits that never appear in a two-qubit gate, and single-qubit
gates in general, carry no entangling structure and so are absent from the
hypergraph; the pass routes them by ownership, assigning each qubit to the
block holding the majority of its two-qubit gates and letting every local gate
follow its qubit. The result is written back as $r$ fragment submodules in
which each severed wire carries the measurement and state-preparation boundary
operations a wire cut requires.

\paragraph{Materialization and lowering} Fragments become executable in two
additional steps within the same task. Let
$\mathcal{A}=\{(I,\lvert0\rangle,\tfrac12),(I,\lvert1\rangle,\tfrac12)\}
\cup\{(P,\lvert P_s\rangle,\tfrac{s}{2}):P\in\{X,Y,Z\},
s\in\{+1,-1\}\}$ denote the eight-element set of measurement, preparation,
and coefficient triples at one cut. For $k$ cuts, term materialization selects
one element of $\mathcal{A}$ per cut, so a complete boundary assignment is
$\alpha=(\alpha_1,\ldots,\alpha_k)\in\mathcal{A}^{\times k}$, where
$\mathcal{A}^{\times k}=\underbrace{\mathcal{A}\times\cdots\times\mathcal{A}}_{k\text{ factors}}$
is the Cartesian product. Its reconstruction coefficient $c_\alpha$ is the product of the selected per-cut coefficients. Each cut contributes eight basis terms, so the task instantiates one concrete fragment experiment per combination and their number grows as $8^k$ before any pruning. It is the cut count, not the size of the circuit, that makes reconstruction expensive, which is what Eq. \eqref{eq:km1} minimizes. Lowering then translates the \texttt{qcut} and quantum-dialect operations of each materialized fragment progressively through the LLVM dialect into the Quantum Intermediate Representation~\cite{QIR2021,qirspec}. The task emits self-contained QIR fragment programs together with an execution manifest recording, for each program, the fragment it belongs to, the experiment it realizes, the measurement basis it uses, and the reconstruction coefficient $c_\alpha$ that the weave task will ultimately apply. This manifest serves as the contract between the cut task and the two downstream stages: it determines how many execute tasks the runtime instantiates and how the weave task combines their results.

\begin{algorithm}[!ht]
  \small
  \caption{QWeaver end-to-end workflow: cut, execute, weave}
  \label{alg:qcut}
  \begin{algorithmic}[1]
    \Require OpenQASM~3 circuit $C$; QPU capacity $Q$; shots $S$; observable $O$
    \Ensure Estimate of $\langle O \rangle$ recovered from fragment samples

    \Statex \textbf{Cut task} \Comment{one task; the only compiler stage}
    \State $M \gets \textsc{RaiseToQuantumDialect}(C)$
    \For{$r \gets 1, 2, \dots$}
        \Comment{smallest $r$ whose fragments fit $Q$}
      \State $(\Pi, \mathcal{C}, \rho) \gets \textsc{Partition}(M, r, \varepsilon, s)$
      \State $F \gets \textsc{Fragments}(M, \Pi, \mathcal{C}, \rho)$
        \Comment{insert measure/prepare boundaries}
      \State \textbf{if} $\max_{f \in F} \mathrm{qubits}(f) \le Q$ \textbf{then break}
    \EndFor
    \State $k \gets |\mathcal{C}|$;\quad
      $\mathcal{A} \gets \textsc{Pauli8Terms}()$
      \Comment{the 8 $(M,\sigma,c)$ terms defined above}
    \State $\mathcal{M} \gets \emptyset$ \Comment{execution manifest}
    \For{each $\alpha \in \mathcal{A}^{\times k}$} \Comment{$8^{k}$ experiments}
      \State $c_{\alpha} \gets \prod_{e=1}^{k} c(\alpha_e)$
      \For{each fragment $f \in F$}
        \State stamp the $(M,\sigma)$ choices of $\alpha$ on the boundary ops of $f$
        \State $p_{\alpha,f} \gets \textsc{LowerToQIR}(f)$
        \State $\mathcal{M} \gets \mathcal{M} \cup
               \{(\alpha, f, p_{\alpha,f}, c_{\alpha}, \mathrm{readout}(f))\}$
      \EndFor
    \EndFor

    \Statex \textbf{Execute tasks} \Comment{$|\mathcal{M}| = r 8^{k}$ tasks, no mutual dependencies}
    \ForAll{$(\alpha, f, p, \cdot, \cdot) \in \mathcal{M}$ \textbf{in parallel}}
      \State $h_{\alpha,f} \gets \textsc{RunQIR}(p, \mathrm{backend}(f), S)$
        \Comment{histogram over $S$ shots}
    \EndFor

    \Statex \textbf{Weave task} \Comment{one task; depends on all execute tasks}
    \State $\hat{P} \gets 0$
    \For{each experiment $\alpha$}
      \For{each fragment $f \in F$}
        \State $g_{\alpha,f}(y) \gets \sum_{z} (-1)^{|z|}\, h_{\alpha,f}(y, z)$
          \Comment{$z$: non-identity boundary bits}
      \EndFor
      \State $P_{\alpha} \gets \prod_{f \in F} g_{\alpha,f}$
             aligned by source qubit id
      \State $\hat{P} \gets \hat{P} + c_{\alpha} P_{\alpha}$
        \Comment{signed quasi-probability}
    \EndFor
    \State \Return $\textsc{Evaluate}(O, \hat{P})$
      \Comment{$\textsc{Project}(\hat{P})$ gives the distribution output}

    \Statex
    \Procedure{Partition}{$M, r, \varepsilon, s$}
      \State trace each qubit value in $M$ to a canonical id
      \State $V \gets \{\,g \in M : g$ is a two-qubit gate on distinct qubits$\,\}$
      \State $E \gets \{\, e_q = \{ g \in V : q \in \mathrm{qubits}(g) \} \,\}$
             \Comment{one hyperedge per qubit}
      \If{$|V| < r$}
        \State $\Pi \gets$ all vertices in block $0$ \Comment{no balanced cut exists}
      \Else
        \State $\Pi \gets \textsc{KaHyPar}(V, E;\ \texttt{km1}, r, \varepsilon, s)$
      \EndIf
      \State $\mathcal{C} \gets \{(g_i, g_j, q)$ : $g_i, g_j$ consecutive on $q$,
             $\Pi(g_i) \neq \Pi(g_j)\}$
      \State $\rho(q) \gets$ block holding the majority of $q$'s two-qubit gates
      \State \Return $\Pi, \mathcal{C}, \rho$
    \EndProcedure
  \end{algorithmic}
\end{algorithm}

\subsection{The execute tasks: QIR fragment execution on QPUs}
\label{subsec:execute-task}

Each entry in the manifest gives rise to one execute task, whose
responsibility is to run a single QIR fragment program on a designated
quantum backend for the requested number of shots and to return the
resulting measurement outcome. Because the execute tasks share no state and
depend only on the cut task, the runtime is free to dispatch them
concurrently across whatever backends are available. This independence is
the source of the framework's parallelism, and it scales jointly with the
number of fragments and the number of quantum processors.

Fragment execution is mediated by a QIR-EE device driver that presents the
QIR-EE execution engine~\cite{wong2025qiree,qireerepo} to the runtime as
another IRIS device. When an execute task is scheduled, the driver loads the
fragment's QIR program from runtime memory, compiles it, determines the
storage required for its expectation values, executes it on the backend named
by the task, and returns the results into the Q-IRIS heterogeneous-memory
model, where they become available to the host as ordinary numerical arrays.
The choice of backend is made independently for each task: fragments may be
simulated directly on Google's QSIM state-vector
simulator~\cite{qsim} or routed through the XACC framework~\cite{xaccrepo} to
reach additional simulators and physical devices such as IonQ trapped-ion
processors~\cite{ionq}. Since the QIR-EE runtime is not thread-safe, the
driver isolates each engine instance in a separate process and exchanges the
QIR program, control information, and returned expectation values with it
through a shared memory-mapped region; this isolation is what allows the
runtime to execute many fragment programs simultaneously where a single
shared engine could not. Partitioning a circuit in this manner reduces the
simulation burden carried by any individual task by more than the number of
fragments, while increasing the number of independently schedulable quantum
tasks, the trade-off that enables early-stage quantum processors to make
progress on circuits they cannot execute as a whole.

\subsection{The weave task: generic post-processing and reconstruction}
\label{subsec:weave-task}

The weave task closes the graph. It has an incoming dependency from every execute task, so Q-IRIS schedules it only after all required fragment experiments have returned and their results are available to the host. The task is therefore entirely classical, its inputs being probability distributions or mean values, together with the reconstruction manifest emitted by the cut task. The manifest is the contract between cutting and reconstruction. For every term $\alpha$, it identifies the coefficient $c_\alpha$, the required location of the fragment-result, the measurement and preparation settings at each boundary, the output-qubit mapping, and the requested target type. The weave kernel consequently does not need knowledge of the original gate sequence.

Both supported target types are instances of the same indexed contraction. Let $y$ denote the retained output index and let $R_{f,\alpha} (y_f)$ be the value supplied by fragment $f$ for the local part $y_f$ of that index.  The weave kernel computes
\begin{equation}\label{eq:reconstruction}
    \hat{R}(y) = \sum_{\alpha \in \mathcal{A}^{\times k}} c_\alpha \prod_{f=1}^r R_{f, \alpha} (y_f) .
\end{equation}

Let $\mu_{f, \alpha}$ be the local expectation value returned by fragment $f$ under boundary assignment $\alpha$. 
For a scalar observable (a mean value), no output index is retained,
so $y=\varnothing$ and
$R_{f,\alpha}(\varnothing)=\mu_{f,\alpha}$. The complete contraction then yields
\[
\hat R(\varnothing)
=
\sum_{\alpha\in\mathcal A^{\times k}}
c_\alpha\prod_{f=1}^{r}\mu_{f,\alpha}
=
\widehat{\langle O\rangle}.
\]
The kernel enumerates the $8^k$ manifest rows, retrieves one local mean $\mu_{f, \alpha}$ from every fragment, forms their product, multiplies by $c_\alpha$, and adds the signed contribution to the accumulator. The quantum processors (physical or simulated) supply the local means; all products and sums are host-side operations. This indexed evaluation is also the operational meaning of  the tensor-network description: each fragment result can be viewed as a small tensor indexed by its boundary settings and measured outcomes, and the weave task contracts the fragment tensors over the common assignment $\alpha$. The implementation does not need to construct a dense tensor, much less an array of dimension $2^n$ (it only reads the entries named by the manifest). 

For a full output distribution $\hat{p}$, the retained output index is a bitstring $x$ and $R_{f, \alpha} (x_{f}) = P_f (x_f | \alpha)$ (the probability assigned by fragment $f$ to its local output $x_f$ when executed under boundary assignment $\alpha$), so that $\hat{R}(x) = \hat{p}(x)$. The manifest maps each local bitstring $x_f$ to its global qubit positions and marks boundary measurements that must be marginalized (trace-out). Sparse local distributions are merged only when their overlapping assignments are compatible, and the resulting signed contribution is accumulated under the corresponding global bitstring.  With finite-shot data, $\hat{p}$ is a quasi-probability estimate and may contain small negative entries or have total mass different from one. QWeaver preserves this signed estimate for diagnostics; when an ordinary probability distribution is requested, it may additionally return a nonnegative normalized projection.  That projection is an output policy, not part of the exact channel identity.  

The generic scalar contraction requires multiplications and accumulations of $O (r 8^k)$ per observable. For a linear chain of fragments, where $r = k + 1$, this is $O(8^k (k+1))$ classical work; it is not another quantum experimental factor. Terms, observables, and output bitstrings can be processed in parallel, provided that their partial sums are reduced consistently. Thus the minimal-cut objective of Section \ref{subsec:cut-task} benefits both sides of the workflow, it means fewer cuts reduce the number of fragment executions and the number of host-side contraction terms.  The weave output is an estimate of the uncut circuit result assembled from fragments that never ran together on one quantum device. 

\section{Cost model}
\label{sec:cost-model}

Circuit cutting exchanges circuit width for additional execution and reconstruction work: narrower fragments can satisfy the backend qubit limit, but every cut expands the quasi-probability decomposition and the task graph that QWeaver must execute. This section develops a first-order model for deciding when that exchange is useful. Throughout, $n$ denotes the original circuit width, $n^{\star}$ the maximum fragment width, $k$ the number of wire cuts, $r$ the number of fragments, and $S$ the shots acquired per materialized fragment program; $m_{Q}$ counts the quantum or simulator workers executing fragments and $m_{C}$ the CPU workers used for reconstruction. Using the GHZ circuit as a tractable case, so its linear CNOT cascade makes every quantity explicit, and we relate circuit width and backend capacity to the number of fragments and cuts, calibrate fragment execution to the selected backend, and separate backend execution work from reconstruction and scheduling. Two distinctions organize the analysis: between holding shots per fragment fixed and holding statistical precision fixed, which differ by a factor of $2^{k}$ in acquisition cost, and between relative speedup over uncut execution and absolute wall-time feasibility under a budget. The objective is not an exact runtime prediction from circuit width alone, but a configuration model that identifies the competing terms, their asymptotic regimes, and the quantities that must be measured. Section~\ref{subsec:cut-count-optima} evaluates these predictions with the end-to-end GHZ experiments.

\subsection{Fragment geometry and decomposition size}
After the initial Hadamard, an $n$-qubit GHZ circuit contains $n-1$
consecutive controlled-$X$ gates. A fragment of width $n^\star$ can contain at
most $n^\star-1$ such gates, so
\begin{equation}
  r(n,n^\star)
    = \left\lceil \frac{n-1}{n^\star-1} \right\rceil,
  \qquad
  k(n,n^\star) = r(n,n^\star)-1.
  \label{eq:ghz-fragment-count}
\end{equation}
The complete Pauli-8 plan contains
\begin{equation}
  L(n,n^\star)=8^{k(n,n^\star)}
  \label{eq:ghz-qpd-count}
\end{equation}
rows and $N_{\mathrm{frag}}=r8^k$ fragment programs. These quantities should
not be conflated: one plan row identifies a boundary assignment and requires
one result from each of the $r$ fragments. The reconstructed mean is
\begin{equation}
  \mu
    = \sum_{\ell=1}^{8^k}c_\ell
      \prod_{j=1}^{r}\mu_{j,\ell},
  \label{eq:ghz-reconstructed-mean}
\end{equation}
where $c_\ell$ is a quasi-probability coefficient and $\mu_{j,\ell}$ the
measured mean of fragment $j$ in row $\ell$. For example, $n=15$ and
$n^\star=2$ require $r=14$ fragments, $k=13$ cuts, and
$8^{13}\simeq5.50\times10^{11}$ plan rows before shots are considered.

\subsection{Calibrated fragment execution}
For a state-vector simulator, a first-order work model for $g$ gates acting on
$q$ qubits is $W(q,g)=g2^q$. Neglecting the single Hadamard, the GHZ cascade
therefore has work $(q-1)2^q$. Let $\tau_0(S)$ be the measured time for an
$n_0$-qubit reference cascade executed with the same shot configuration as the
benchmark. The time of a width-$q$ fragment is approximated by
\begin{equation}
  t(q,S)
    = \tau_0(S)
      \frac{(q-1)2^q}{(n_0-1)2^{n_0}}.
  \label{eq:ghz-calibrated-time}
\end{equation}
Writing the calibration as $\tau_0(S)$ is important: a physical QPU normally
pays for repeated shots, whereas a state-vector simulator may evolve the state
once and sample it many times. The measured calibration absorbs that backend
dependence instead of assuming that every simulator cost is linear in $S$.
The corresponding uncut execution estimate is
\begin{equation}
  \tau_f^{\mathrm{QC}}(n,S)
    = \tau_0(S)
      \frac{(n-1)2^n}{(n_0-1)2^{n_0}}.
  \label{eq:ghz-full-time}
\end{equation}

\subsection{Parallel execution and classical reconstruction}
The $r8^k$ fragment programs are mutually independent once the cut task has
emitted the manifest. Approximating every fragment by the limiting width
$n^\star$, a global QWeaver schedule across $m_Q$ homogeneous workers has the
idealized execution time
\begin{equation}
  \tau_{c,\mathrm{global}}^{\mathrm{QC}}
    = \left\lceil\frac{r8^k}{m_Q}\right\rceil t(n^\star,S).
  \label{eq:ghz-cut-time-global}
\end{equation}
A more conservative schedule completes all fragments for one plan row before
starting the next, giving
\begin{equation}
  \tau_{c,\mathrm{sync}}^{\mathrm{QC}}
    = 8^k\left\lceil\frac{r}{m_Q}\right\rceil t(n^\star,S).
  \label{eq:ghz-cut-time}
\end{equation}
Within this equal-task approximation, a real schedule lies between these two
estimates before runtime overheads; fragment imbalance, batching, and device
placement determine where. Neither expression includes the classical weave
task. For $N_O$ independently contracted observables, a simple parallel
reconstruction model is
\begin{equation}
  \tau_c^{\mathrm{PM}}
    \simeq N_O\left\lceil\frac{8^k}{m_C}\right\rceil
      \left[(r-1)t_\times+t_+\right]+t_{\mathrm{red}},
  \label{eq:ghz-postmeasurement-time}
\end{equation}
where $t_\times$ and $t_+$ are measured scalar product and accumulation times,
and $t_{\mathrm{red}}$ is the final parallel-reduction cost. The complete model
is consequently
\begin{equation}
  \tau_c^{\mathrm{total}}
    = \tau_c^{\mathrm{QC}}+\tau_c^{\mathrm{PM}}
      +\tau_{\mathrm{sched}}+\tau_{\mathrm{I/O}}.
  \label{eq:ghz-total-time}
\end{equation}
This separation matters in practice: the $r$ factor counts fragment executions,
whereas the weave task contracts $8^k$ products; it does not create another
factor of $r$ in the number of plan rows.

Ignoring the overhead terms and using the synchronized estimate, division by
Eq.~\eqref{eq:ghz-full-time} gives
\begin{equation}
  R_{\mathrm{sync}}^{\mathrm{QC}}
    \equiv \frac{\tau_{c,\mathrm{sync}}^{\mathrm{QC}}}
                   {\tau_f^{\mathrm{QC}}}
    = 8^k\left\lceil\frac{k+1}{m_Q}\right\rceil
      \frac{n^\star-1}{n-1}2^{n^\star-n}.
  \label{eq:ghz-time-ratio}
\end{equation}
The factor $2^{n^\star-n}$ is the state-vector benefit of narrower fragments;
$8^k$ is the decomposition penalty; and the ceiling is the part that task
parallelism can reduce. A value below one predicts only that fragment execution
is faster than the extrapolated uncut simulation. End-to-end break-even instead
requires $\tau_c^{\mathrm{total}}<\tau_f^{\mathrm{total}}$, where the latter is
measured on the same backend and includes the corresponding overheads.

\subsection{Fixed shots, fixed precision, and absolute feasibility}
Figure~\ref{fig:ghz-cut-scaling} holds $S=1024$ fixed, so its dominant
cut-dependent acquisition factor is $8^k$. Holding statistical precision fixed
is more expensive. Equation~\eqref{eq:harris-ross-shot-overhead} gives
$S_{\mathrm{eff}}=S/2^k$ for the complete Pauli-8 scheme; maintaining a constant
effective shot count therefore requires $S(k)=S_0 2^k$. Including the $r$
fragment results per plan row, the acquisition work scales as
\begin{equation}
  \mathcal W_{\mathrm{acq}}(k)\propto
  \begin{cases}
    rS_0 8^k,  & \text{fixed shots per fragment},\\
    rS_0 16^k, & \text{fixed statistical precision}.
  \end{cases}
  \label{eq:ghz-fixed-precision-work}
\end{equation}
The second line is a sampling-work statement. It translates directly into wall
time for shot-limited hardware, but a simulator that reuses one evolved state
for many samples can have weaker shot dependence; this is why
Eq.~\eqref{eq:ghz-calibrated-time} uses a measured $\tau_0(S)$.

A relative speedup is not by itself a feasibility result, since both the cut and
uncut times may be impractically large. Given an execution allocation or
wall-time limit $T_{\mathrm{budget}}$, the operational condition is
\begin{equation}
  \tau_c^{\mathrm{total}}(n,n^\star,m_Q,m_C,S)
    \leq T_{\mathrm{budget}}.
  \label{eq:ghz-absolute-feasibility}
\end{equation}
Thus a claim that an $n$-qubit circuit can be run with an $n^\star$-qubit
backend is incomplete unless it also reports the cut count, full plan size,
shots or target uncertainty, concurrency, measured fragment throughput, and
wall-time budget.

\subsection{Asymptotic interpretation}
For balanced fragments,
$n^\star-1=(n-1)/(k+1)+O(1)$. With a fixed worker count, the leading base-two
exponent of the synchronized acquisition time is
\begin{equation}
  \log_2\tau_c^{\mathrm{QC}}
  =
  \begin{cases}
    3k+\dfrac{n}{k+1}+O(\log n), & \text{fixed shots},\\[6pt]
    4k+\dfrac{n}{k+1}+O(\log n), & \text{fixed precision}.
  \end{cases}
  \label{eq:ghz-log-time}
\end{equation}
Treating $k$ as continuous gives the idealized optima
\begin{equation}
  k_{\mathrm{opt}}\simeq
  \begin{cases}
    \sqrt{n/3}-1, & \text{fixed shots},\\
    \sqrt{n}/2-1, & \text{fixed precision},
  \end{cases}
  \qquad
  \tau_{c,\mathrm{opt}}^{\mathrm{QC}}=
  \begin{cases}
    2^{2\sqrt{3n}+O(\log n)},\\
    2^{4\sqrt n+O(\log n)}.
  \end{cases}
  \label{eq:ghz-optimal-cuts}
\end{equation}
These subexponential expressions assume that the fragment width is allowed to
grow as $\Theta(\sqrt n)$ and therefore do not describe a fixed-capacity QPU.
If $n^\star$ is fixed, then $k=\Theta(n/(n^\star-1))$ and
\begin{equation}
  \tau_c^{\mathrm{QC}}
    = 2^{a n/(n^\star-1)+O(\log n)},
  \qquad
  a=
  \begin{cases}
    3, & \text{fixed shots},\\
    4, & \text{fixed precision}.
  \end{cases}
  \label{eq:ghz-fixed-budget-asymptotic}
\end{equation}
Relative to uncut state-vector work $\Theta(n2^n)$, the exponential rate
improves only for $n^\star>4$ at fixed shots and $n^\star>5$ at fixed
precision. Very narrow fragments can therefore lose their width advantage to
the decomposition and sampling overheads.

\begin{figure*}[t]
  \centering
  \includegraphics[width=\textwidth]{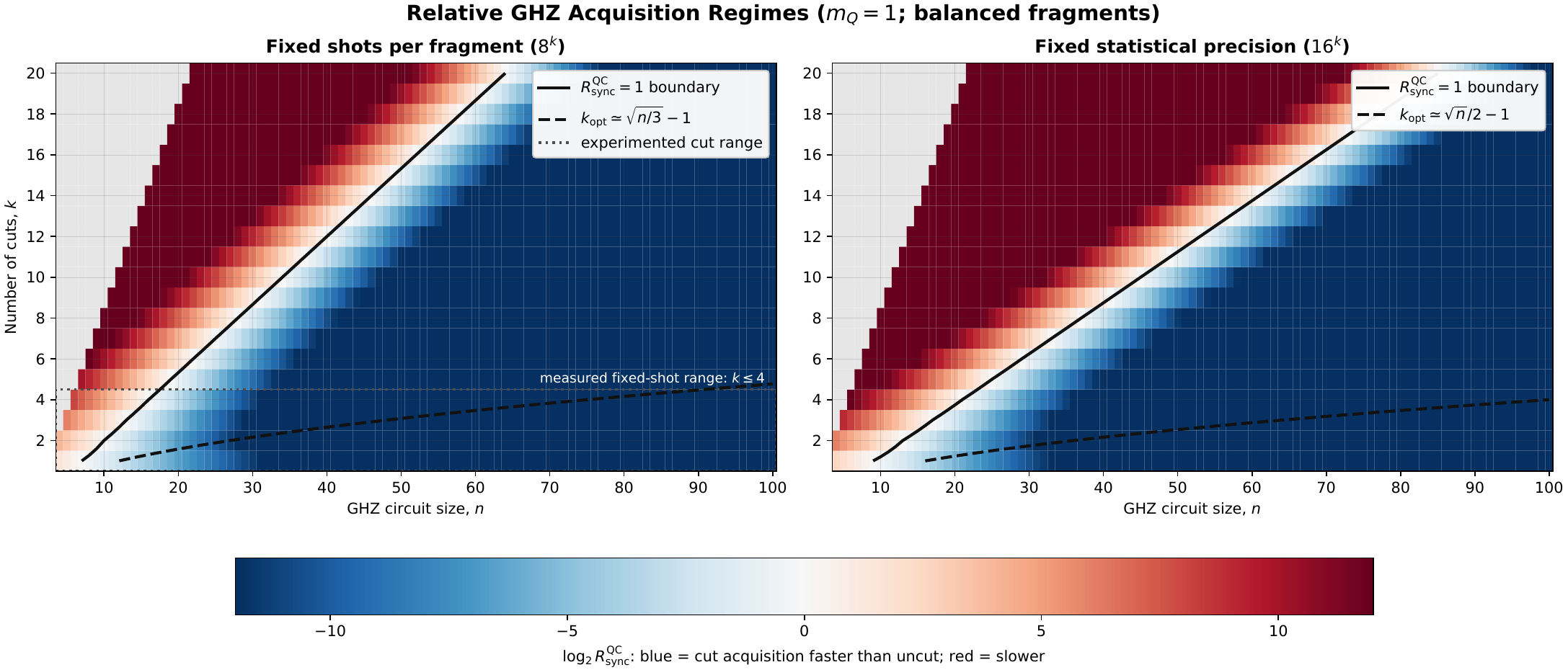}
  \caption{Asymptotic relative acquisition regimes for balanced GHZ fragments
  with one quantum or simulator worker ($m_Q=1$). Color gives
  $\log_2 R_{\mathrm{sync}}^{\mathrm{QC}}$ for fixed shots per fragment (left)
  and fixed statistical precision (right): blue predicts that cut-fragment
  acquisition is faster than the extrapolated uncut simulation, red predicts
  that it is slower, and the solid black curve is the break-even boundary
  $R_{\mathrm{sync}}^{\mathrm{QC}}=1$. Maintaining precision adds the
  $2^k$ shot factor to the $8^k$ complete-plan cost, producing a $16^k$
  acquisition penalty and shifting the break-even frontier toward larger
  circuits. The dashed curves show the corresponding leading-order optima,
  $k_{\mathrm{opt}}\simeq\sqrt{n/3}-1$ and
  $k_{\mathrm{opt}}\simeq\sqrt n/2-1$. The dotted box marks the fixed-shot
  cut range explored in Fig.~\ref{fig:ghz-cut-scaling}, and gray cells are
  infeasible ($k>n-2$). This comparison covers idealized quantum acquisition
  only; it does not include reconstruction, scheduling, or I/O overheads and
  does not by itself establish the absolute wall-time condition in
  Eq.~\eqref{eq:ghz-absolute-feasibility}.}
  \label{fig:ghz-time-model-regions}
\end{figure*}

This model yields two testable predictions. First, from the fixed-shot branch: each additional cut lowers the slope of the exponent in $n$ but raises its intercept by a factor of eight, so the preferred cut count should remain moderate and grow slowly with circuit size, rather than additional cuts being uniformly beneficial. For $n=60$, for example, Eq.~\eqref{eq:ghz-optimal-cuts} gives $k_{\mathrm{opt}}\simeq3.5$, making three and four cuts the leading integer candidates; the experiments of Fig.~\ref{fig:ghz-cut-scaling} satisfy the width assumption behind Eq.~\eqref{eq:ghz-optimal-cuts}, since the backend capacity is adjusted to each cut count. Second, from the cost separation of Eq.~\eqref{eq:ghz-total-time}: a relative acquisition speedup need not imply end-to-end feasibility, because scheduling, materialization, and reconstruction enter the total independently of acquisition. The experiments in Section~\ref{subsec:cut-count-optima} test both predictions, the first through the measured time-minimizing cut counts and the second through the wall-time timeout boundary.

\paragraph{Overhead of the chosen decomposition}
\label{par:overhead-alternatives}%
QWeaver materializes the complete eight-term Pauli decomposition, a per-cut factor of eight and an $O(16^{k})$ reconstruction sampling overhead for $k$ cuts. This is not the minimal achievable overhead. Allowing classical communication between fragments lowers the optimal $k$-cut cost from $O(16^{k})$ to $O(4^{k})$~\cite{Piveteau2024,Brenner2023}, and randomized measurements reduce the constant factor without such communication~\cite{Lowe2023}. We deliberately retain the communication-free decomposition because it keeps fragment experiments mutually independent, which is precisely what lets the runtime dispatch them concurrently across heterogeneous and physically separated backends; the communication-assisted variants would require mid-circuit measurement and cross-device feed-forward, reintroducing dependency edges among the execute tasks. Reductions that need no inter-fragment communication, such as randomized-measurement estimators, are directly compatible with the task graph and are a natural optimization of the cut task's term materialization, which we leave to future work.

\section{Experimental results}
\label{sec:results}

\subsection{GHZ benchmark results}
\label{subsec:results-integration}

\begin{figure}[!ht]
  \centering
  \includegraphics[width=0.6\linewidth]{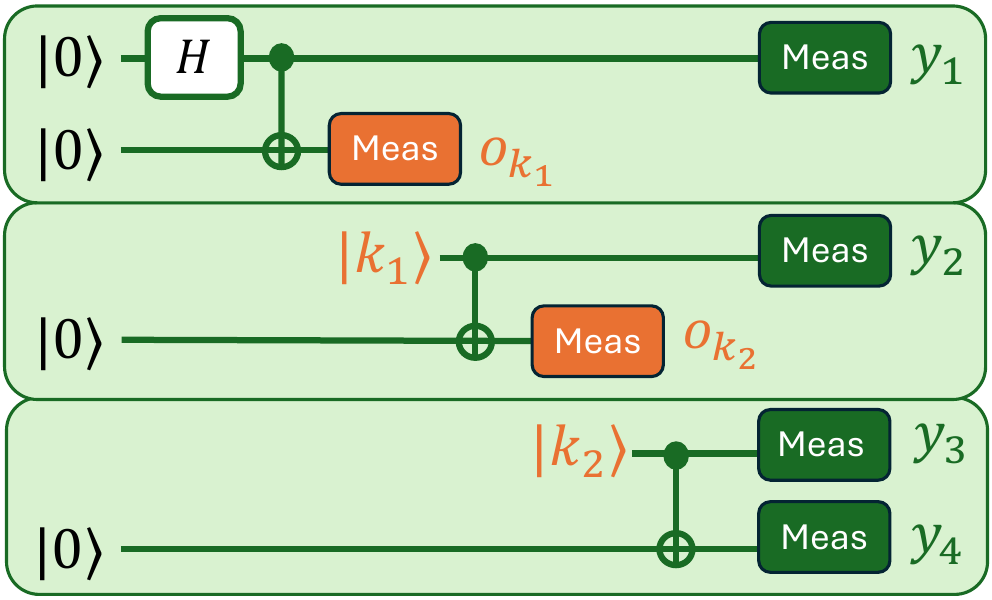}
  \caption{Partitioning a 4-qubit GHZ circuit with two wire cuts into
    three two-qubit subcircuits. For each QPD experiment, the upstream boundary measurements (orange) and corresponding preparations $|k_1\rangle$ and $|k_2\rangle$ are fixed when the fragment programs are
    materialized and the fragments are executed independently. The weave task combines the resulting boundary values $o_{k_1},o_{k_2}$ and fragment outputs $y_1,\dots,y_4$ with the signed QPD coefficients to estimate $\langle ZZZZ\rangle$~\cite{Peng2020,harada2024doubly,miniskar2026q}.}
  \label{fig:qcut-fig-simple}
  \vspace{-0.2in}
\end{figure}

To demonstrate the complete workflow, we evaluate QWeaver on the four qubit
Greenberger-Horne-Zeilinger (GHZ) circuit. The two wire cuts produce three two-qubit fragments, as illustrated in
Fig.~\ref{fig:qcut-fig-simple}. Each cut contributes eight fixed
measure-and-prepare terms, so the complete plan contains $8^2=64$ QPD
experiments. Term materialization, described in
Section~\ref{subsec:cut-task}, emits one program per fragment and experiment,
for $192$ fragment programs in total~\cite{miniskar2026q,qcutrepo}. All
measurement and preparation choices are encoded in those programs before
dispatch, so the three programs in an experiment have no runtime feed-forward
dependency.

This workflow extends the earlier Q-IRIS/Q-CUT prototype from an execution
experiment into an end-to-end compiler/runtime pipeline. At its user-facing
entry point, QWeaver accepts an OpenQASM~3 circuit and the cut task raises it
into the MLIR quantum dialect. Circuit cutting, QIR lowering and package
generation, fragment execution, and reconstruction are then represented as
dependent tasks in one graph.
Consequently, compilation is part of the measured workflow rather than an
external preparation step. Each fragment program executes through the QIR-EE
driver of Section~\ref{subsec:execute-task}, reaching Google's QSIM state
simulator directly or additional simulators through XACC. Once all execute
tasks complete, the weave task aggregates their outcomes and evaluates the
QPD reconstruction of Eq.~\eqref{eq:reconstruction} to estimate
$\langle ZZZZ\rangle$, whose ideal value is one.

Table~\ref{tab:qweaver-ghz004-backends} reports a fresh execution of
this workflow using $1024$ shots per fragment and eight Q-IRIS virtual QPU
slots where applicable. For the simulator rows, we time three sequential
stages: MLIR cutting and QIR package generation (Compile), execution of all
$192$ fragment programs (Runtime), and quasi-probability evaluation
(Reconstruct). The IQM workflow expresses the same operations as one dependent
task graph; its three stage values are task-group spans, while Total is the
complete wall time and includes orchestration outside those spans. The
benchmark harness retains per-task scheduler timings and all emitted
reconstruction diagnostics.

\begin{table*}[!ht]
\centering
\caption{QWeaver backend comparison for the 4-qubit GHZ circuit ($1024$
  shots per fragment, two cuts, $64$ QPD experiments, $192$ fragment
  programs, and eight local Q-IRIS virtual QPU slots where applicable). The
  accuracy baseline is the uncut, noiseless GHZ result
  $\langle ZZZZ\rangle_{\mathrm{ideal}}=1$.
  Error is the observed absolute deviation from the ideal value
  (the mean over ten workflows for IQM), not a propagated standard error or
  confidence interval. The simulator rows test whether cutting,
  execution, and reconstruction preserve a known observable.
  The Q-IRIS/IQM row summarizes ten complete routed workflows in
  three time-separated submission sessions, each with pre/post readout calibration.
  The session-level uncertainty and readout sensitivity are reported below.
}
\label{tab:qweaver-ghz004-backends}
\setlength{\tabcolsep}{5pt}
\renewcommand{\arraystretch}{1.1}
\resizebox{\textwidth}{!}{%
\begin{tabular}{lrrrrrr}
\hline
\textbf{Backend}
& \boldmath$\langle ZZZZ \rangle$
& \textbf{Error}
& \textbf{Compile (s)}
& \textbf{Runtime (s)}
& \textbf{Reconstruct (s)}
& \textbf{Total (s)}\\
\hline\hline
QIR-QSIM             & 1.000000 &  0.00\% & 1.1130 &   2.9886 & 0.1620 &   4.2635\\
XACC/QPP             & 0.960841 &  3.92\% & 1.1149 &   3.7231 & 0.1618 &   4.9998\\
XACC/QSIM            & 1.011280 &  1.13\% & 1.1356 &   3.7316 & 0.1611 &   5.0284\\
XACC/AER             & 1.029403 &  2.94\% & 1.1212 &   3.7663 & 0.1640 &   5.0515\\
XACC/IONQ            & 1.000000 &  0.00\% & 1.1214 &  76.9762 & 0.1602 &  78.2578\\
XACC/IONQ:SIM.ARIA-1 & 0.943898 &  5.61\% & 1.1246 & 171.9291 & 0.1652 & 173.2190\\
\hline
Q-IRIS/IQM (10 runs, mean) & 0.715236 & 28.48\% & 3.8269 & 93.1468 & 0.08230 & 102.8902\\
\hline
\end{tabular}%
}
\end{table*}

Across the six simulator configurations, compilation and reconstruction are
nearly backend-independent:
compilation ranges from $1.11$ to $1.14$ seconds and reconstruction from
$0.16$ to $0.17$ seconds. Backend execution therefore determines the
difference in end-to-end time. Direct QIR-QSIM is fastest at $2.9886$ seconds,
while the XACC local simulators require $3.7231$--$3.7663$ seconds. The IonQ
ideal emulator and noisy Aria-1 simulator~\cite{ionq} take $76.9762$ and
$171.9291$ seconds, respectively. These IonQ tasks are submitted through the
vendor cloud service, so their runtime includes client/API communication,
cloud submission, and service-side scheduling overhead in addition to
simulation. The compile and reconstruction phases remain unchanged, showing
that QWeaver isolates this backend-specific cost within the execute tasks.

\label{par:iqm-hardware-runs}%
Our hardware experiments consisted of ten independent complete
workflows across three submission sessions for each of GHZ-4 and GHZ-6.
Every workflow used the same routed IQM Pathfinder backend, six-qubit physical region (i.e., qubits 3,4,5,8,9,10 on IQM's Crystal topology), two-cut 64-experiment/192-fragment plan, and 1024 shots
per fragment.  With the complete workflow as the experimental unit, raw
GHZ-4 reconstructions have mean 0.7152, sample SD 0.0461, and a 10,000 sample 95\% interval $[0.6868,0.7402]$; GHZ-6 gives 0.6426,
0.0351, and $[0.6228,0.6635]$, respectively.  Their mean absolute deviations
from the ideal value are 0.2848 and 0.3574.

These measurements used a single routed physical-QPU endpoint and therefore do not constitute simultaneous execution across multiple physical QPUs. The parallel experiments reported elsewhere use virtual QPU slots or independent simulator instances; consequently, the present evaluation does not measure physical-QPU speedup, parallel efficiency, or cross-QPU scheduling overhead. A controlled evaluation using multiple concurrently available and comparably calibrated physical QPUs is left for future work. More generally, these hardware results are limited to GHZ circuits on a single routed backend and qubit region; they do not establish performance for other circuit families or devices.

\begin{figure}[!ht]
  \centering
  \includegraphics[width=0.9\linewidth]{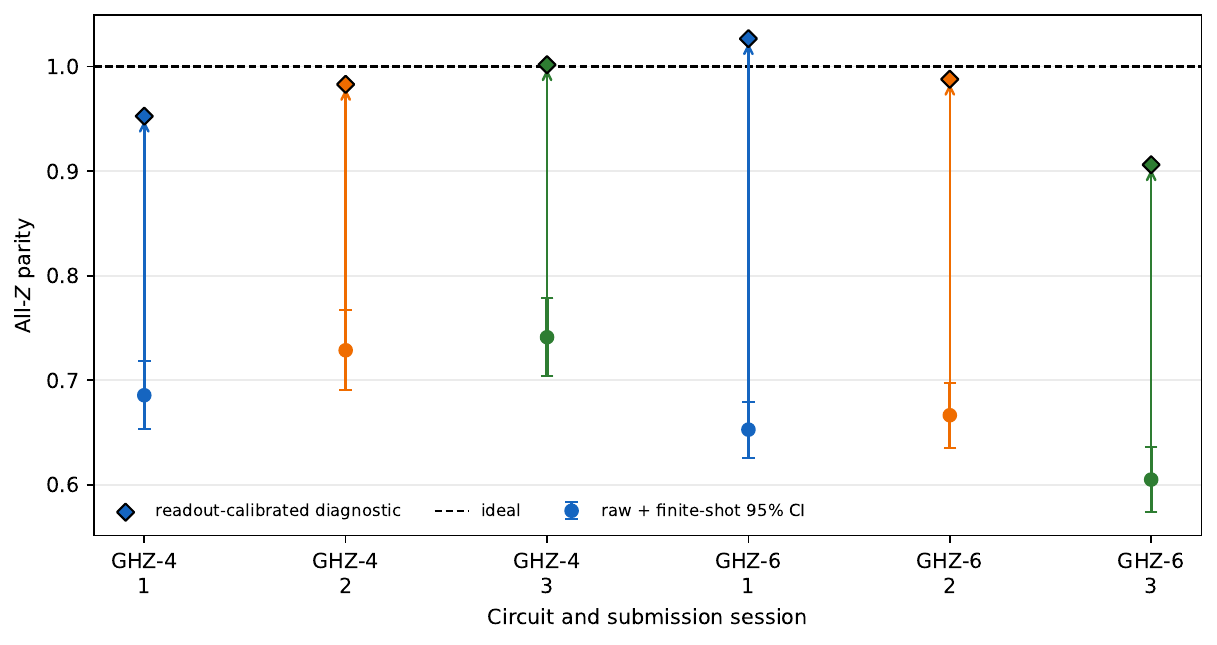}
  \caption{Raw IQM reconstruction (circles) with conditional
  finite-shot 95\% intervals for each circuit/session mean.  Arrows terminate
  at the secondary readout-calibrated diagnostics (diamonds); the dashed line
  is the ideal value one.  The raw values and their intervals are the primary
  results.}
  \label{fig:iqm-uncertainty-budget}
\end{figure}

Each hardware session included readout-calibration measurements
before and after the GHZ workloads, together with mapping-matched uncut
controls.  We used these measurements to estimate how sensitive the
reconstructed observable is to readout error.  Both the fragment counts and
calibration counts were resampled 10,000 times when computing the reported
intervals in Table~\ref{tab:iqm-uncertainty-budget} and Fig.~\ref{fig:iqm-uncertainty-budget}.

\begin{table*}[!ht]
\centering
\caption{IQM uncertainty and readout-sensitivity diagnostics by
  circuit and submission session.  The finite-shot interval is the two-sided
  percentile 95\% interval for the raw session mean.  The readout shift is
  the readout-calibrated diagnostic minus the raw result; its interval jointly
  resamples fragment and calibration counts.  Raw results are the primary
  hardware measurements.}
\label{tab:iqm-uncertainty-budget}
\setlength{\tabcolsep}{3.5pt}
\renewcommand{\arraystretch}{1.1}
\resizebox{\textwidth}{!}{%
\begin{tabular}{llrrrrr}
\hline
\textbf{Circuit} & \textbf{Session} & \boldmath$\mathbf{n}$ & \textbf{Raw}
& \textbf{Raw finite-shot 95\% CI} & \textbf{Readout-calibrated diagnostic}
& \textbf{Readout shift (95\% CI)}\\
\hline\hline
GHZ-4 & S1 & 4 & 0.6857 & $[0.6530,0.7181]$ & 0.9525 & $+0.2668\ [0.2465,0.2876]$\\
GHZ-4 & S2 & 3 & 0.7287 & $[0.6905,0.7669]$ & 0.9829 & $+0.2542\ [0.2341,0.2750]$\\
GHZ-4 & S3 & 3 & 0.7412 & $[0.7042,0.7784]$ & 1.0017 & $+0.2605\ [0.2405,0.2819]$\\
GHZ-6 & S1 & 4 & 0.6528 & $[0.6259,0.6793]$ & 1.0266 & $+0.3738\ [0.3520,0.3962]$\\
GHZ-6 & S2 & 3 & 0.6666 & $[0.6351,0.6978]$ & 0.9878 & $+0.3212\ [0.3001,0.3424]$\\
GHZ-6 & S3 & 3 & 0.6051 & $[0.5742,0.6364]$ & 0.9061 & $+0.3010\ [0.2806,0.3221]$\\
\hline
\end{tabular}%
}
\end{table*}

\label{par:iqm-uncertainty-budget}%
The raw results and propagated finite-shot intervals are our main
hardware findings.  As a sensitivity check, applying the readout calibration
changes the overall GHZ-4 mean from 0.7152 to 0.9764 and the GHZ-6 mean from
0.6426 to 0.9788.  We report these adjusted values only to show the possible
scale of the readout contribution and they are not used to claim a new mitigation method or improved hardware accuracy. Because QPD reconstruction combines fragment results with signed weights, it can amplify finite-shot fluctuations. The reported hardware intervals include this effect.

\label{par:error-mitigation}%
We did not apply error mitigation in any of the experiments reported here. The hardware values are raw reconstructions, and the readout calibration of Table~\ref{tab:iqm-uncertainty-budget} and Fig.~\ref{fig:iqm-uncertainty-budget} is reported only as a sensitivity diagnostic, not as a correction applied to the results. We instead outline how mitigation would enter the workflow, as additional tasks rather than as a change to the cut identity. A confusion-matrix readout correction would be a task inserted between each execute task and the weave task, and a gate-level scheme such as randomized compiling would appear as a compilation option on the cut task or as a wrapper on the execute tasks. Because the eight-term QPD sum combines fragment estimates with signed coefficients, any residual readout or gate error is amplified during reconstruction, so mitigation is most effective applied per fragment, before the signed sum. A calibrated, task-graph-integrated mitigation stage, together with its own propagated uncertainty, is left to future work.

All six simulator configurations in Table~\ref{tab:qweaver-ghz004-backends}
complete the same $192$-task manifest and reconstruct $\langle ZZZZ\rangle$
within $0.0562$ of its ideal value.  The QIR-QSIM value in that table motivates
the seeded validation below and its exact value in Table~\ref{tab:qweaver-ghz004-backends}
should not be interpreted as an absence of finite-shot error. 
Peng's eight-term QPD identity is algebraically exact, but it amplifies error
in estimated fragment probabilities.  The XACC/IONQ ideal cloud simulator is
different: it returns deterministic counts equal to the ideal probabilities
scaled by $1024$, leaving no independent finite-shot residual for the signed
QPD sum to amplify.

We corrected the QIR-QSIM path by deriving each fragment seed from
the tuple (base seed, trial ID, QPD experiment ID, fragment ID) and ran 20 fresh,
complete-plan GHZ-4 trials for QIR-QSIM (with XACC/QPP and XACC/AER for comparison)
using $1024$  shots per fragment, 64 QPD experiments, and 192 fragment programs per trial. 
The 3,840 derived QIR-QSIM fragment seeds  were distinct across the 20 trials.

\begin{figure}[!ht]
  \centering
  \includegraphics[width=0.78\linewidth]{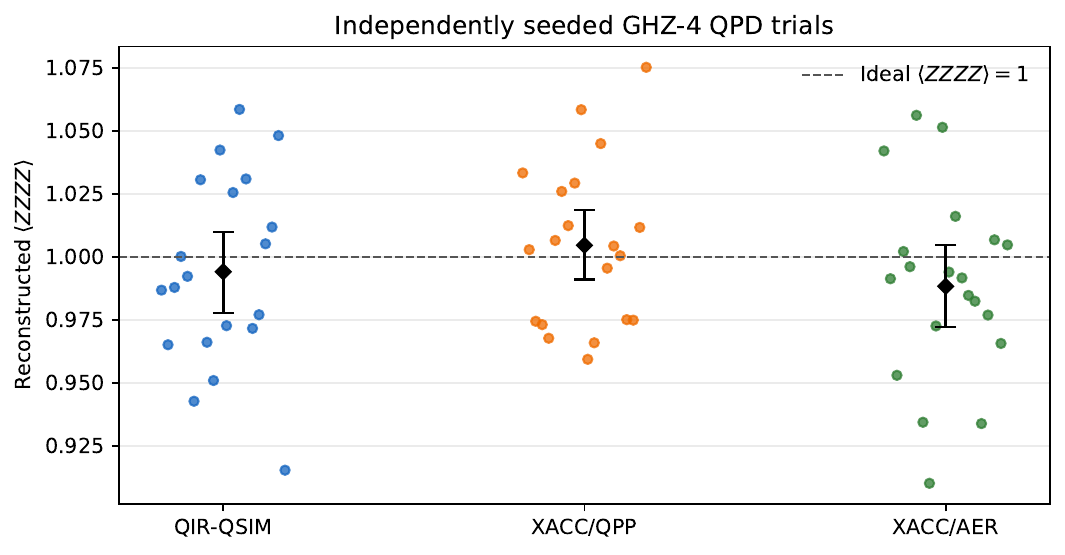}
  \caption{Trial-level reconstructed $\langle ZZZZ\rangle$.
  Points are the 20 independent complete-plan trials per
  backend; diamonds and error bars show the mean and 95\% confidence interval
  from 10,000 resamples. The dashed line is the ideal value one.}
  \label{fig:seed-validation}
\end{figure}

As Fig.~\ref{fig:seed-validation} shows, QIR-QSIM exhibits
finite-shot variation ($0.9154$--$1.0586$).
Its mean is $0.9941$ with sample standard deviation
$0.0379$ and 95\% confidence interval $[0.9778,1.0099]$.
XACC/QPP gives $1.0046\pm0.0326$ with interval $[0.9910,1.0187]$, and
XACC/AER gives $0.9883\pm0.0379$ with interval $[0.9721,1.0046]$, where the
reported plus/minus values are sample standard deviations across trials.
All three intervals include the ideal value one.

Taken together, the experiment validates QWeaver as an end-to-end improvement
over the earlier Q-IRIS/Q-CUT workflow. The cut task performs MLIR cutting,
QIR lowering, and package generation inside the task graph; Q-IRIS then
executes the resulting fragments through direct and XACC-backed simulators or
the routed IQM hardware backend; and the weave task reconstructs the target
observable. Compilation is
therefore embedded as part of the workflow rather than being treated as a 
separate prerequisite to runtime execution.

\subsection{Harris--Ross railway problem}
\label{subsec:harris-ross}

The GHZ experiments establish the correctness of the complete QWeaver workflow,
but their linear interaction graph makes the placement of a cut nearly trivial.
We therefore use the Harris--Ross railway network as a structured
combinatorial-optimization benchmark in which the partitioner must contend with
an irregular, weighted topology. Harris and Ross originally studied the
capacity of a railway network spanning the western Soviet Union and Eastern
Europe~\cite{HarrisRoss1955}; the report subsequently became an important
historical precursor to the maximum-flow/minimum-cut problem~\cite{Schrijver2002}.
Our digitized fixture removes auxiliary boundary and exit information and
contains 43 usable railway divisions connected by 94 weighted links, with
weights derived from the reported capacities. As an independent transcription
check, the fixture gives a maximum flow of 164 (in thousands of tons per day)
from divisions $\{8,10\}$ to $\{1,4\}$, whereas the historical source reports
163. We retain 164 as the reproducible value of the digitized fixture and do
not treat the one-unit difference as a correction to the historical result.

The quantum workload is weighted MaxCut, rather than the classically
polynomial flow problem. For $G=(V,E)$ with edge capacities $w_{uv}$, the
measured objective is
\begin{equation}
  \langle C\rangle
    = \sum_{(u,v)\in E}
        \frac{w_{uv}}{2}\left(1-\langle Z_u Z_v\rangle\right).
  \label{eq:harris-ross-maxcut}
\end{equation}
We encode this objective as a depth-one QAOA
circuit~\cite{Farhi2014}: one logical qubit represents each division, the
initial state is $|+\rangle^{\otimes 43}$, and each railway link contributes the
cost-phase operator
$\exp[-\mathrm{i}\gamma w_{uv}Z_uZ_v/2]$. Because the QIR execution path used
in these experiments does not provide a native weighted $R_{ZZ}$ operation,
the compiler emits this term as
\textsc{cnot}--$R_Z(\gamma w_{uv})$--\textsc{cnot}; a terminal
$R_X(\beta)$ is applied to every qubit. The resulting circuit contains 368
operations: $43$ Hadamards, three operations for each of the 94 weighted links,
and $43$ mixer rotations. We fix $\gamma=0.25$ and $\beta=0.15$ throughout.
The fixed angles deliberately remove classical parameter optimization from the
experiment, isolating circuit partitioning, fragment execution, and
reconstruction; consequently, the results below are not claims about QAOA
solution quality.

\paragraph{Complete-plan validation}
An exact reconstruction of the full 43-qubit circuit is too costly to use as a
correctness test. We instead form a validation ladder of induced subgraphs with
$n=8,10,12,$ and $14$ qubits by breadth-first expansion from the documented
origin divisions. These instances retain the local structure and weights of the
railway network while keeping both the complete Pauli-8 decomposition and an
independent state-vector calculation tractable. For induced subgraphs with $n>14$, a two- or three-cut partition does not reduce the largest fragment below the original width, because the cut boundary qubits are counted in each fragment. Useful reductions require about five cuts at $n=18$ and $20$ and six at $n=22$, corresponding to $8^5=32,768$ and $8^6=262,144$ plan rows, which exceeds the roughly $65,000$ fragment executions already needed for the $14$-qubit validation. Every cut contributes eight
measure-and-prepare terms, and for each ladder instance we execute all $8^k$
experiment rows for $k$ wire cuts, using 4096 shots per fragment on the
Q-IRIS/XACC QPP state-vector backend. We verified from the emitted manifests
that every experiment contains exactly one term for every cut and that the
experiment tuples form the complete Cartesian product.

\begin{table}[!ht]
  \centering
  \caption{Harris--Ross validation ladder using complete Pauli-8
    decompositions and 4096 shots per fragment. Relative error is
    $|\langle\widehat C\rangle-\langle C\rangle|/W$, where
    $W=\sum_{(u,v)\in E}w_{uv}$ is the total edge weight.}
  \label{tab:harris-ross-validation}
  \setlength{\tabcolsep}{9pt}
  \begin{tabular}{rrrrrrr}
    \toprule
    Qubits & Edges & Cuts $k$ & Widths
      & Exact & Reconstructed & Error \\
    \midrule
     8 & 12 & 2 & $5,5$ & 42.7286 & 42.6894 & 0.04\% \\
    10 & 17 & 2 & $5,7$ & 60.3748 & 59.5412 & 0.69\% \\
    12 & 21 & 4 & $5,9$ & 74.5520 & 74.8666 & 0.21\% \\
    14 & 24 & 5 & $8,9$ & 88.2078 & 87.6442 & 0.32\% \\
    \bottomrule
  \end{tabular}
\end{table}

\begin{figure*}[t]
  \centering
  \includegraphics[width=\textwidth]{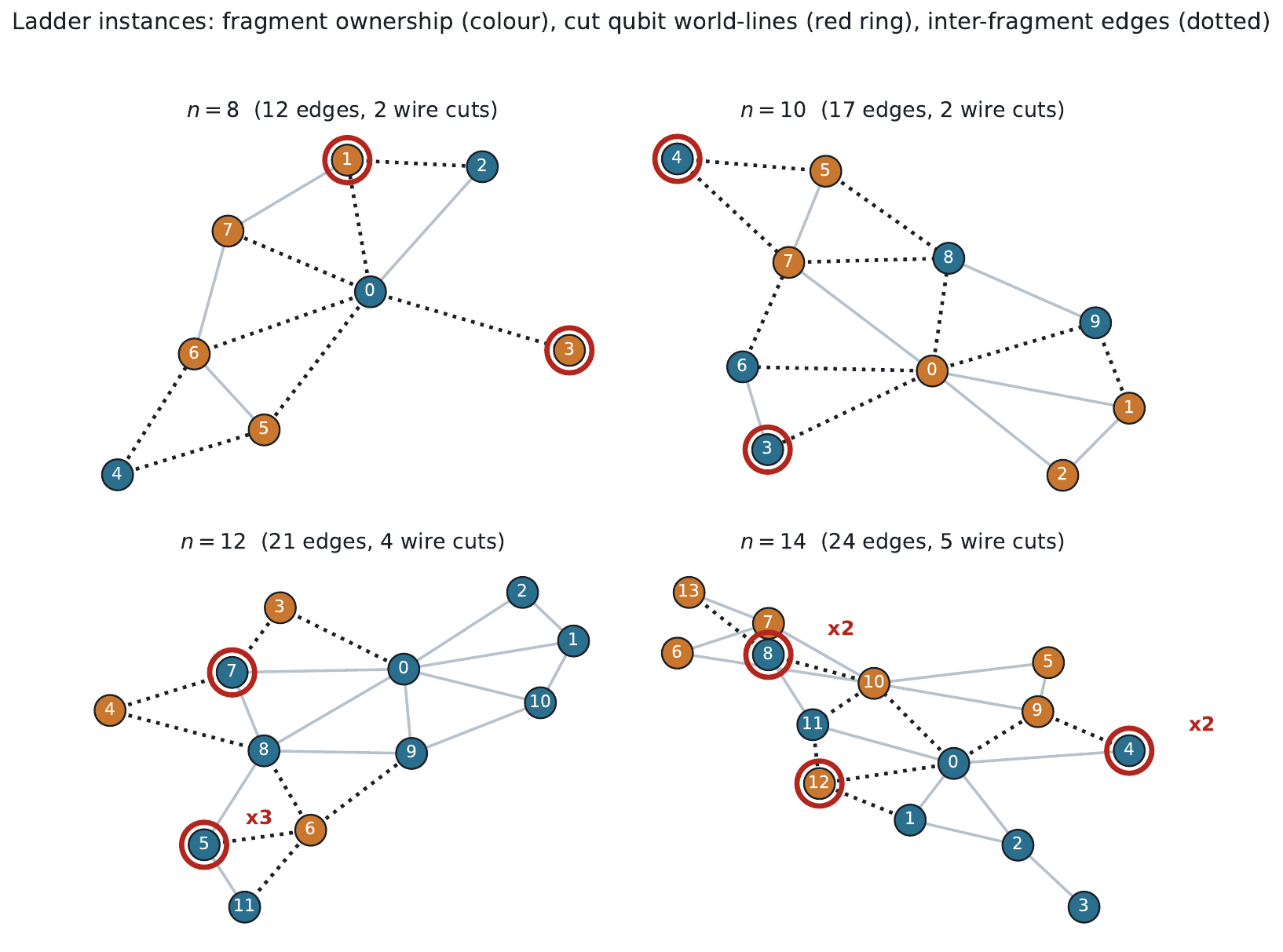}
  \caption{Partitions of the Harris--Ross validation ladder. Node color denotes
    fragment ownership, dotted links join qubits assigned across fragment
    boundaries, and red rings identify severed qubit worldlines; a multiplicity
    is shown when the same worldline is cut more than once. The four instances
    require two, two, four, and five wire cuts, respectively.}
  \label{fig:harris-ross-partitions}
\end{figure*}

Figure~\ref{fig:harris-ross-partitions} shows the partitions and temporal wire
cuts selected for this ladder. A dotted problem-graph edge joins qubits whose
gates reside in different fragments, whereas a red ring marks a qubit worldline
that is actually severed. The two counts need not agree: QWeaver partitions
ordered gate occurrences and inserts a wire cut only where ownership changes
between consecutive entangling operations on a worldline, as described in
Section~\ref{subsec:cut-task}.

Table~\ref{tab:harris-ross-validation} shows that the reconstructed cut value
agrees with exact simulation to $0.04$--$0.69\%$ of the total edge weight across
the complete-plan instances. The observed deviations are consistent with
finite-shot error. Let $\widehat{\mu}_\alpha$ denote the complete
product of fragment estimates for row $\alpha$, and assume that different rows
are estimated independently with comparable variance
$\operatorname{Var}(\widehat{\mu}_\alpha)=\sigma^2/S$, where $S$ is the
number of shots per fragment program. The emitted coefficients have uniform
magnitude $|c_\alpha|=2^{-k}$ over $8^k$ rows and satisfy
$\sum_\alpha c_\alpha^2=2^k$. For $S$ shots per fragment, the corresponding
variance and effective shot count are therefore
\begin{equation}
  \operatorname{Var}\!\left(\widehat{\langle O\rangle}\right)
    = \sum_\alpha c_\alpha^2
      \operatorname{Var}(\widehat{\mu}_\alpha)
    \simeq \frac{2^k\sigma^2}{S},
  S_{\mathrm{eff}}=\frac{S}{2^k}.
  \label{eq:harris-ross-shot-overhead}
\end{equation}
Holding the statistical error fixed requires $S\propto2^k$ shots for each of
the $8^k$ rows, giving total sampling work proportional to $16^k$. This
measured cost law makes the partitioner's cut objective a systems concern as
well as a compiler concern.

Observable agreement does not imply that the complete output distribution is
equally accurate. Across the same ladder, total variation distance from the
exact distribution rises from 0.161 and 0.199 at two cuts to 0.475 and 0.661 at
four and five cuts. Moreover, the fixed depth-one state places the expected cut
value within approximately one percent of the uniform-sampling baseline. The
optimum bitstring appears in the sampled support, but is not consistently
enriched relative to uniform sampling. We therefore use
Eq.~\eqref{eq:harris-ross-maxcut} as the validation target and interpret these
experiments as evidence for accurate observable reconstruction, not as evidence
of a MaxCut advantage.

\paragraph{Full-instance deployment}
For the 43-qubit circuit, fragment budgets of 35, 30, and 25 qubits all select the same natural bisection: two fragments of 23 and 24 qubits separated by eight wire cuts. QWeaver thus executes a 43-qubit workload while no execute task simulates more than 24 qubits, a width gain of $43/24 = 1.79$. For each budget we materialized 512 of the $8^{8}$ experiment rows, yielding 1,024 fragment-execution tasks. Q-IRIS schedules within a single node, so the multi-node deployment uses the manifest itself as the distribution mechanism: because the rows are mutually independent, the manifest partitions cleanly by experiment identifier into eight shards, each executed by an independent Q-IRIS instance on one node of the OLCF Frontier system under the same configuration as the single-node experiments. This deployment confirmed balanced execution across shards, checkpoint and resume after a wall-time interruption (each manifest row is a self-contained unit of work, so completed rows are skipped on restart), and bit-identical accumulated outputs from the three independently compiled packages. Package construction for the 30- and 25-qubit budgets took approximately one hour in each case.

These full-instance runs demonstrate orchestration throughput, not a converged
observable. The 512 rows comprise only $0.003\%$ of the required
$8^8=16,777,216$-row decomposition, so their partial quasi-probability sums
cannot be assigned a reconstruction error. A complete plan would contain
approximately $3.4\times10^7$ fragment executions and, under the measured
configuration, require approximately $1.1\times10^5$ node-hours before adding
the extra shots that may be required to maintain fixed statistical precision under Eq.~\eqref{eq:harris-ross-shot-overhead}. The
Harris--Ross benchmark therefore exposes both sides of circuit cutting: QWeaver
successfully converts an application circuit beyond the backend width into a
large parallel task graph, while the potential $O(16^k)$ fixed-precision sampling work of the complete Pauli-8 plan places a sharp premium on minimizing the number of wire cuts.

\subsection{Empirical cut-count optima and feasibility}
\label{subsec:cut-count-optima}

\label{par:scalability-scope}%
The GHZ-32 and GHZ-60 circuits in this subsection are shallow,
linear, highly structured simulator workloads used to study systems, width,
and cut-count trade-offs.  They do not represent arbitrary circuits of the
same width, and the observed optima should not be generalized to deeper or
irregular algorithms, to circuits whose required cut count grows with problem
complexity, or to physical hardware. 

\begin{figure*}[!t]
  \centering
  \includegraphics[width=\textwidth]{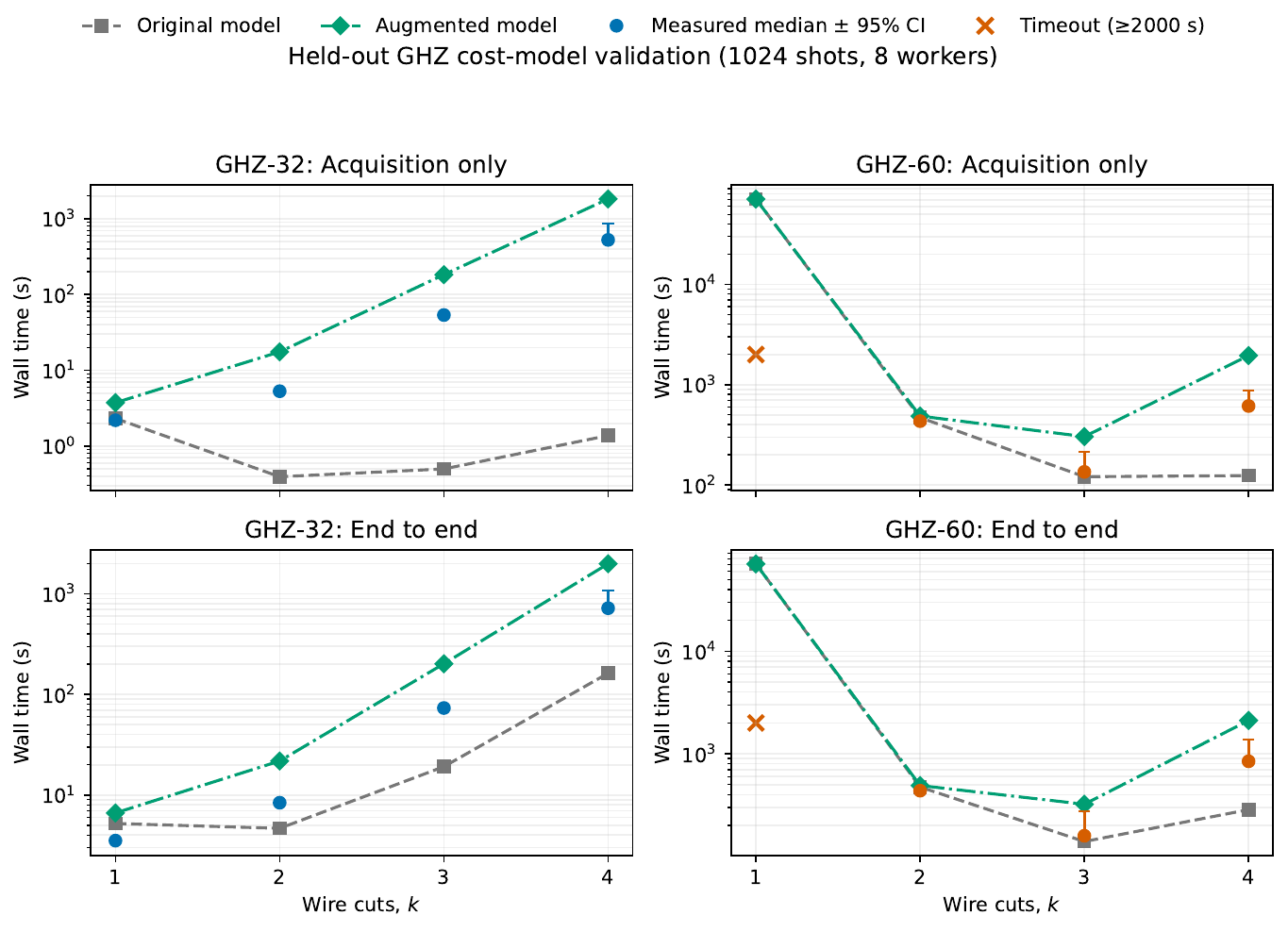}
  \caption{GHZ model validation on the Q-IRIS/XACC QPP
    simulator.  The fragment-time law is calibrated once from five independent
    uncut workflows at each width $12$, $16$, $20$, and $24$, then evaluated at
    the actual fragment widths for GHZ-32 and GHZ-60 with $m_Q=8$ workers and
    $1024$ shots per fragment.  The upper row reports acquisition time; the
    lower row adds compile, scheduling/I/O, and reconstruction
    times.  Points and error bars are measured medians and 
    two-sided percentile 95\% intervals from five workflows per completed
    configuration (10,000 resamples).  Dashed lines show
    the original width-only model, dash-dotted lines show the single global
    per-task-intercept augmentation, and the cross marks the measured
    $2000\,\mathrm{s}$ GHZ-60 one-cut timeout as a lower bound.}
  \label{fig:ghz-cut-scaling}
\end{figure*}

Figure~\ref{fig:ghz-cut-scaling} compares the calibrated cost model
with end-to-end measurements for GHZ-32 and GHZ-60 at one through four
cuts.  For $k$ cuts, each completed workflow executes the full Pauli-8 plan of
$8^k$ rows and $r8^k$ fragment programs, where $r=k+1$, using $1024$ shots per fragment on the Q-IRIS/XACC QPP statevector simulator and eight workers.  Every completed point summarizes five independent workflows.  The GHZ-60 one-cut workflow
reached our prespecified $2000\,\mathrm{s}$ limit and is retained as a censored
lower bound rather than a completed measurement.

These measurements expose the width--work trade-off directly.
For GHZ-32 the one-cut configuration is fastest, whereas for GHZ-60 the
three-cut configuration is fastest.  Additional cuts reduce the actual
fragment widths but multiply the complete Pauli-8 plan by eight, so the optimum
depends on both circuit width and fixed per-task costs.

\paragraph{From predicted trade-offs to observed optima}

The original fragment law, fitted only to the 20 uncut calibration
workflows, gives $\tau_0(20,1024)=11.013\,\mathrm{s}$ with a 95\% interval of
$[10.975,11.050]\,\mathrm{s}$.  Predictions use the actual fragment widths and
a global greedy schedule over eight workers; the end-to-end curves additionally
use global overhead fits from the calibration data.  A single allowed
augmentation adds a nonnegative per-task intercept of
$0.714\,\mathrm{s}$, with interval $[0.638,0.773]\,\mathrm{s}$.

For GHZ-60, the measured median end-to-end times are
$436.85\,\mathrm{s}$ (95\% CI $[436.08,441.74]$) at two cuts,
$158.86\,\mathrm{s}$ ($[157.36,276.26]$) at three cuts, and
$845.27\,\mathrm{s}$ ($[829.82,1382.11]$) at four cuts.  The median
difference $T_3-T_2=-278.00\,\mathrm{s}$ has interval
$[-282.88,-160.59]$, while $T_4-T_3=686.42\,\mathrm{s}$ has interval
$[569.01,1223.25]$; both exclude zero.  Together with the censored
$T_1\geq2000\,\mathrm{s}$ observation, these comparisons support three cuts as
the measured GHZ-60 optimum.  All three adjacent GHZ-32 differences likewise
exclude zero and support its one-cut optimum.

The validation also bounds the model's accuracy rather than only its
ordering.  Across the seven completed configurations, the original
end-to-end prediction-to-measurement ratios range from $0.226$ to $1.480$. In other words, it
selects the GHZ-60 optimum but reverses the GHZ-32 one- versus two-cut ordering.
The augmented model recovers every adjacent ordering, but its ratios range from
$1.123$ to $2.763$ and systematically overpredict the many-task cases.  Thus a
single launch captures calibration-scale fixed cost but not the
batched and amortized scheduler behavior well enough for point prediction.

For the repeated GHZ-24 workflows in Table~\ref{tab:ghz24-cut-ratios}, the median total wall time fell from $214.27\,\mathrm{s}$ (95\% CI $[213.36,215.64]$) without cuts to $1.86\,\mathrm{s}$ ($[1.82,1.91]$) with one cut, giving a speedup $T_0/T_1=115.39$ ($[112.47,117.64]$).
Beyond the one-cut optimum, $T_2/T_1=4.42$ ($[4.04,4.51]$) and $T_3/T_2=8.75$ ($[8.55,9.49]$); both intervals exclude unity.

A cut configuration that is faster than an uncut extrapolation is
useful only if its end-to-end time satisfies the absolute budget in
Eq.~\eqref{eq:ghz-absolute-feasibility}.  The measured timeout therefore
complements the crossover intervals: together they identify both the fastest
feasible cut count and where the width--work exchange ceases to finish under
the experimental limit.

\begin{table*}[t]
  \centering
  \caption{Repeated GHZ-24 end-to-end timings and ratios relative to the uncut workflow. Values are medians over five independent QPP workflows; brackets give trial-level bootstrap 95\% confidence intervals (10,000 resamples). Cuts reduce fragment width, but each additional cut multiplies the complete Pauli-8 plan by eight; the optimum is therefore workload- and backend-dependent rather than uniformly improving with every cut.}
  \label{tab:ghz24-cut-ratios}
  \footnotesize
  \begin{tabular}{@{}cccc@{}}
    \toprule
    Cuts $k$ & $T_k$ (s) & Overhead $T_k/T_0$ & Speedup $T_0/T_k$ \\
    \midrule
    0 & 214.27 [213.36, 215.64] & 1.00 [1.00, 1.00] & 1.00 [1.00, 1.00] \\
    1 & 1.86 [1.82, 1.91] & 0.0087 [0.0085, 0.0089] & 115.39 [112.47, 117.64] \\
    2 & 8.22 [7.57, 8.32] & 0.0383 [0.0353, 0.0388] & 26.08 [25.76, 28.29] \\
    3 & 71.88 [70.21, 72.59] & 0.335 [0.328, 0.339] & 2.98 [2.95, 3.05] \\
    \bottomrule
  \end{tabular}
\end{table*}

\subsection{Fragment-job growth versus per-fragment memory}
\label{subsec:cut-memory-tradeoff}

\begin{figure*}[t]
  \centering
  \includegraphics[width=\textwidth]{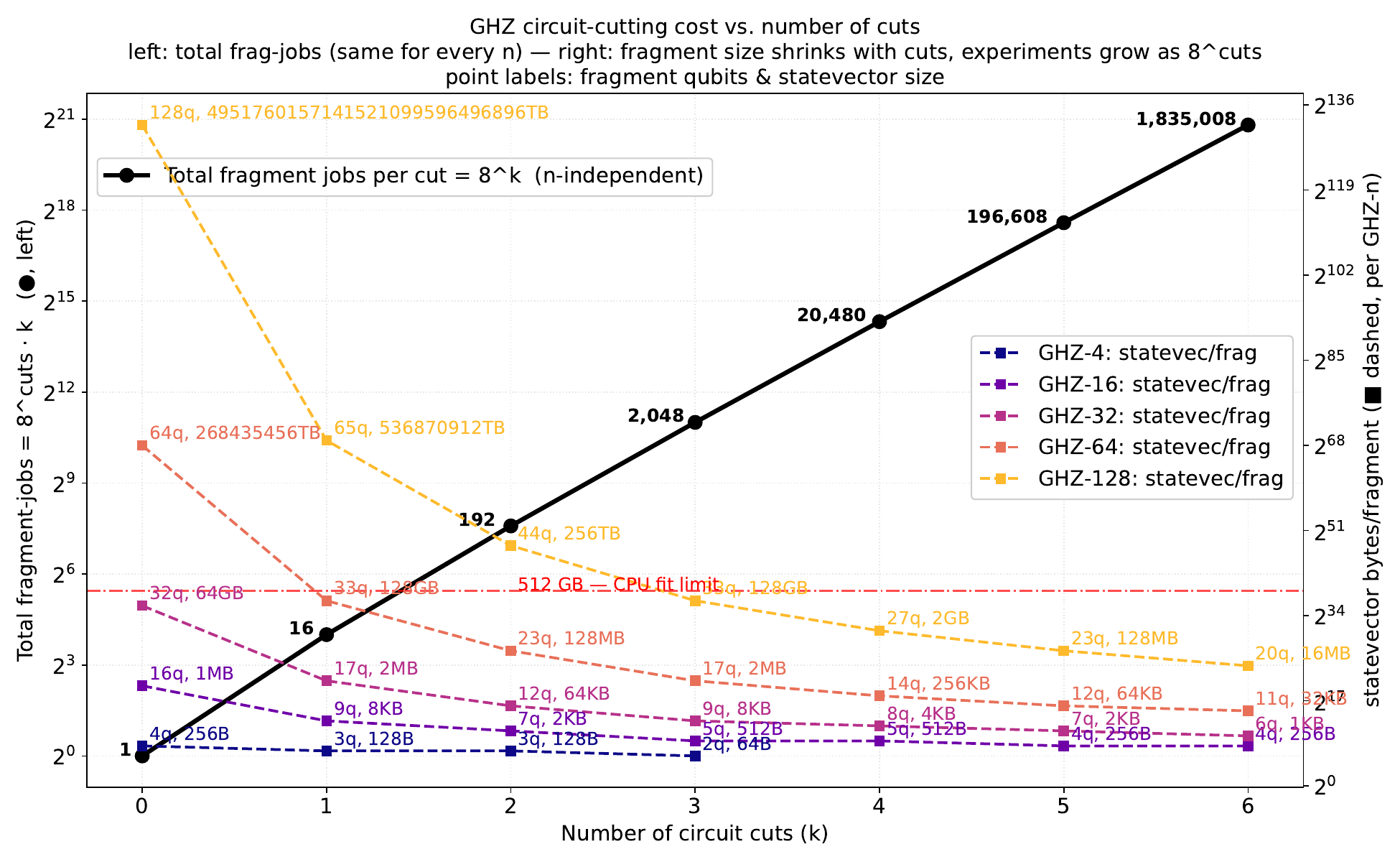}
  \caption[Circuit-cutting cost--memory trade-off for GHZ circuits.]{Analytical circuit-cutting cost--memory projection for GHZ circuits, shown against the
    number of cuts $k=r-1$ on log$_2$ axes. The solid curve on the left axis is
    the total number of fragment jobs $8^{k}(k+1)$, which is independent of the
    circuit-size label $n$ used in the figure and is dominated by the
    factor-of-eight growth per cut. The dashed
    curves on the right axis are the dense-statevector memory per fragment,
    $2^{w}\!\times\!16$ bytes with fragment width $w\approx\lceil n/r\rceil+1$,
    with one curve per $n$, and each point is labelled with its fragment width
    and statevector size. The dash-dotted line marks a $512\,\mathrm{GB}$
    host-memory budget. Additional cuts shrink every fragment below the memory
    line but move all circuits up the shared job curve. Large circuits such as
    GHZ-128 therefore require several cuts before any fragment fits, whereas
    GHZ-32 and smaller circuits already fit at a single cut.}
  \label{fig:ghz-cut-memory}
\end{figure*}

Figure~\ref{fig:ghz-cut-memory} is an analytical resource projection,
not a record of executed circuits.  It quantifies the circuit-cutting trade-off for GHZ
circuits of $n\in\{4,16,32,64,128\}$ qubits as a function of the cut count
$k$. Partitioning a linear chain into $r$ fragments introduces $k=r-1$ cut
edges, and Pauli-8 reconstruction expands each cut into eight measure/prepare
terms. The total number of fragment jobs is $8^{k}r = 8^{k}(k+1)$. This quantity,
plotted as the solid curve in Fig.~\ref{fig:ghz-cut-memory}, is independent of $n$
and is dominated by the factor-of-eight growth per cut, increasing from $16$ at
$k=1$ to $1,835,008$ at $k=6$. The dependence on $n$ is negligible because the
linear factor $r=k+1$ is subdominant to the exponential factor $8^{k}$.

The per-fragment memory requirement decreases as $k$ increases. Each fragment is
represented as a dense statevector of $2^{w}\times 16$ bytes, where the fragment
width is $w\approx\lceil n/r\rceil+1$. For $k=0$ the circuit forms a single
fragment of width $n$, requiring $2^{n}\times 16$ bytes, equal to
$16\,\mathrm{GB}$ at $n=30$, $16\,\mathrm{TB}$ at $n=40$, and
${\sim}5\times10^{39}$ bytes at $n=128$. Because the requirement is exponential
in $w$, small $k$ reduces it only marginally for large $n$: GHZ-128 requires
$65$ qubits ($512\,\mathrm{EB}$) per fragment at $k=1$ and $44$ qubits
($256\,\mathrm{TB}$) at $k=2$. The per-fragment statevector first falls below
$512\,\mathrm{GB}$ for GHZ-128 at $k=3$ ($33$ qubits, $128\,\mathrm{GB}$) and
below $80\,\mathrm{GB}$ at $k=4$ ($27$ qubits, $2\,\mathrm{GB}$), whereas GHZ-32
satisfies the $512\,\mathrm{GB}$ bound at $k=1$ ($2\,\mathrm{MB}$). The minimal
feasible configuration is therefore the smallest $k$ for which $2^{w}\times 16$
does not exceed the device memory. Each additional cut multiplies the number of
plan rows by eight, while the wall-clock effect also depends on the reduction in
fragment width, scheduling, and fixed per-task costs. Circuit cutting thus
exchanges the $O(2^{n})$ per-fragment memory requirement, which limits uncut
single-node simulation to $n\approx35$--$40$, with an $O(8^{k})$ fragment-job
cost; the reconstruction sampling requirement can scale as $O(16^{k})$ without
classical communication.  This analytical exchange does not establish execution
feasibility for arbitrary GHZ-128 circuits or for deeper, irregular workloads.

\section{Related work}
\label{sec:related-work}

Research work related to QWeaver falls into three groups: the quasi-probability theory that determines what a cut costs, the search procedures that decide where a circuit should be cut, and the runtime systems that execute the resulting fragments. QWeaver contributes to the second and third groups, a cut-placement pass and a task-graph runtime for the fragments it produces. Table~\ref{tab:related-work} summarizes the comparison.

\subsection{Decompositions and sampling overhead}

\cite{Peng2020} evaluated large circuits on smaller devices by cutting edges of the circuit's tensor network into clusters and recombining the clusters classically, at a cost exponential in the qubits communicated between them. \cite{Mitarai2021} recast the problem as quasi-probability sampling, writing a non-local channel as a signed combination of local ones whose one-norm $\gamma$ multiplies across cuts and yields a sampling overhead of $\gamma^{2}$. Every later result is a statement about $\gamma$. \cite{Piveteau2024} showed that classical communication between subcircuits, and cutting several gates jointly, both lower it; \cite{Brenner2023} proved that the optimal cost of $k$
wire cuts is $O(16^{k})$ without such communication and $O(4^{k})$ with it; and \cite{Lowe2023} reduce the constant factor using randomized measurements.

\cite{harada2024doubly} minimize a second metric alongside $\gamma$. Their
decomposition of the parallel $n$-wire identity channel uses mutually unbiased
bases and no ancilla qubits, and attains both the minimal sampling overhead and
the minimal channel count $2^{n}+1$. The channel count matters here because
each channel is a separate program to materialize, lower, and schedule, which
is what dominates the cut task of Section~\ref{subsec:cut-task}.

QWeaver does not adopt this joint minimal-channel decomposition.
Instead, it applies the eight-term Pauli measure-and-prepare decomposition
independently to each cut wire, producing $8^k$ boundary assignments for $k$
cuts. Compared with the joint $k$-wire construction of
\cite{harada2024doubly}, this choice materializes more terms, but preserves a
factorized per-cut representation that maps directly to QWeaver's independent
execute tasks and manifest-based reconstruction.

QWeaver dispatches fragments to backends that share no real-time classical
channel, so it forgoes the reduced overhead of the communication-assisted
decompositions~\cite{Piveteau2024,Brenner2023}. This is a property of the
decomposition the cut task materializes, not of the task graph: adopting one
would change the manifest entries and the weave coefficients, and would require
mid-circuit measurement and feed-forward across devices.

\subsection{Cut placement}

CutQC~\cite{Tang2021} selects wire cuts with a mixed-integer program under a
per-subcircuit qubit budget and reconstructs the output classically, reporting
evaluations of circuits up to 100 qubits. The Circuit Knitting Toolbox cut
finder~\cite{Shehzad2024} searches the gate sequence best-first, allowing each
gate to be left intact, cut, or cut on either or both of its wires, ordered so
that the first complete solution minimizes $\gamma^{2}$.
\cite{Froehler2026} admit gate cuts, wire cuts, and joint cuts of gate groups
in one objective, and reduce placement to balanced graph partitioning because
the exact formulations do not scale beyond roughly 40 qubits. The FindCut
procedure of Qdislib~\cite{Tejedor2025arxiv,Tejedor2025} partitions a graph
derived from the circuit with Kernighan-Lin, Girvan-Newman, spectral, and METIS
solvers, choosing among their results by a weighted loss over cut count,
subcircuit count, and qubits per subcircuit.

Adaptive circuit knitting~\cite{Johnson2026ack} places cuts by entanglement
instead. Since the overhead of a bipartition is governed by the robustness of
entanglement across it, a low-entanglement boundary is a low-overhead boundary.
The method represents the state as a tensor network, optimizes a circuit per
partition variationally, and relocates cuts using entanglement-entropy
heatmaps, reducing overheads by up to four orders of magnitude on disordered
mixed-field Ising models of up to 60 qubits.

Our objective is weaker and cheaper. Scoring a boundary by entanglement targets
the quantity that actually determines $\gamma$, but it requires a variational
model of the state, so placement depends on gate parameters and on an
optimization whose difficulty grows with the entanglement it measures. QWeaver
minimizes the number of cut wires, a structural property recovered from the IR
in one pass, and therefore treats all cut wires as equally expensive. The two
are complementary: an entanglement estimate enters our formulation as the
hyperedge weights $\omega(e)$ of Eq.~\eqref{eq:km1}. Our model also differs
from the graph formulations above in that a qubit wire is a single hyperedge
however many gates it threads, so \texttt{km1} charges it once per additional
block it visits. That is a lower bound on the cuts the extraction step inserts,
since a wire that leaves a block and re-enters is cut twice, and it is what
multilevel hypergraph partitioners minimize at scale.

\subsection{Frameworks for cutting and hybrid execution}

Qdislib~\cite{Tejedor2025,Tejedor2025arxiv} is closest to QWeaver in scope. It
cuts a gate-level DAG derived from a Qiskit or Qibo circuit and executes the
subcircuits through PyCOMPSs across CPUs, GPUs, and QPUs, with all cutting
logic inside the library and PyCOMPSs used only to parallelize execution.
DQR~\cite{Raigada2026} argues for the opposite separation and decouples
cutting from orchestration through an immutable fragment descriptor, so its
coordinator can dispatch fragments in waves, retry them, and reroute them from
a QPU to a simulator without reading circuit code.
Pilot-Quantum~\cite{Mantha2025} applies the pilot abstraction to quantum
resources and treats circuit cutting as one workload pattern among several.

In all three the cut happens before the schedule and outside it: a library
produces fragments, and an orchestration layer places them. QWeaver admits the
cut into the schedule. The cut task is a node of the same graph as the execute
and weave tasks, its manifest determines how many execute tasks the runtime
instantiates, and its outputs reach them through the Q-IRIS
heterogeneous-memory model rather than through staged files. The fan-out width
is therefore data-dependent without external coordination, and fragments are
lowered to QIR inside the cut task and run by QIR-EE~\cite{wong2025qiree} as
another IRIS device, so one artifact serves simulators and hardware alike.

\subsection{Compiler infrastructure for quantum programs}

Several quantum compilation stacks build on MLIR: a dialect for quantum
assembly languages~\cite{McCaskey2021}, QSSA~\cite{Peduri2022}, which enforces
the physical constraints of qubit use through value semantics,
QIRO~\cite{Ittah2022}, which pairs memory-semantics and value-semantics
dialects for co-optimization, and Catalyst~\cite{Ittah2024Catalyst}. These
lower to QIR~\cite{QIR2021,qirspec}.

We are not aware of a circuit-cutting transformation implemented as a pass over
any of these dialects. The cut finders above operate on Python circuit objects
or graphs derived from them, which leaves partitioning outside the compilation
flow. Performing it in MLIR lets the wire tracing reuse the def-use chains
already in the IR, and produces fragments in the input dialect, so the existing
lowering path to QIR applies unchanged.

\begin{table*}[t]
\centering
\caption{Circuit-cutting frameworks and cut-placement methods compared with
  QWeaver. The final column indicates whether the work provides an execution
  layer for the fragments it produces.}
\label{tab:related-work}
\footnotesize
\setlength{\tabcolsep}{6pt}
\renewcommand{\arraystretch}{1.1}
\resizebox{\textwidth}{!}{%
\begin{tabular}{@{}lllll@{}}
\toprule
\textbf{Work} & \textbf{Placement method} & \textbf{Cut type} &
\textbf{Representation} & \textbf{Fragment execution} \\
\midrule
CutQC~\cite{Tang2021}           & Mixed-integer program     & Wire
  & Python circuit object   & Small QPUs, postprocessing \\
CKT cut finder~\cite{Shehzad2024} & Best-first search on $\gamma^{2}$ & Gate and wire
  & Qiskit circuit          & Qiskit primitives \\
Gate groups~\cite{Froehler2026} & Balanced graph partitioning & Gate, wire, groups
  & Gate graph              & Placement only \\
ACK~\cite{Johnson2026ack}       & Entanglement heatmaps     & Gate
  & Tensor network          & GPU simulators, QPUs \\
Qdislib~\cite{Tejedor2025,Tejedor2025arxiv} & FindCut graph partitioners & Gate and wire
  & Gate DAG in Python      & PyCOMPSs tasks \\
DQR~\cite{Raigada2026}          & Delegated to cutter       & Inherited
  & Fragment descriptor     & Wave-based dispatch \\
Pilot-Quantum~\cite{Mantha2025} & Delegated to cutter       & Inherited
  & Qiskit, PennyLane       & Pilot-managed tasks \\
\midrule
QWeaver (this work)              & Hypergraph partitioning   & Wire
  & MLIR quantum dialect    & Q-IRIS tasks via QIR-EE \\
\bottomrule
\end{tabular}}
\end{table*}

\section{Conclusion}
\label{sec:conclusion}

We presented QWeaver, a task-graph framework that expresses the entire quantum
circuit-cutting pipeline of circuit cutting, fragment execution, and result
reconstruction as a single directed acyclic graph of first-class tasks
scheduled by the Q-IRIS heterogeneous runtime. In this graph, an MLIR-based cut
task raises an OpenQASM~3 circuit into the quantum and \texttt{qcut} dialects,
partitions its entangling structure with a hypergraph model and KaHyPar, and
emits QIR fragment programs that the execute tasks run through the QIR-EE engine
and the weave task recombines by quasi-probability reconstruction. Its principal
advance over our earlier Q-IRIS/Q-CUT prototype~\cite{miniskar2026q} is that
cutting is now a compiler transformation on a typed, value-semantic intermediate
representation rather than an external, hand-prepared step. This yields three
concrete benefits. First, entanglement analysis, partitioning, term
materialization, and QIR lowering compose as verified passes within one
intermediate representation. Second, compilation is part of the measured
end-to-end workflow and is nearly backend-independent, taking $1.11$--$1.14$~s
against $0.16$--$0.17$~s for reconstruction, so the end-to-end time is dominated
by fragment execution. Third, because fragments are emitted as standard QIR, a
single cut result is portable across heterogeneous backends, including Google
QSIM directly and XACC-routed QPP, Aer, and IonQ, and it composes with the
broader MLIR quantum ecosystem~\cite{Ittah2024Catalyst,qecompiler,cudaq}.
Empirically, the reconstructed $\langle Z^{\otimes n}\rangle$ for
GHZ circuits agrees with its ideal value to within $0.0562$ across six simulator backends.
Our width and cut-count measurements use shallow, linear,
structured GHZ simulator workloads and characterize QWeaver's systems
behavior for such cases and the analytical study in
Section~\ref{subsec:cut-memory-tradeoff} shows the governing exchange:
reducing the $O(2^{n})$ per-fragment statevector memory requires an
$O(8^{k})$ fragment-job plan and can entail $O(16^{k})$ reconstruction sampling
overhead for $k$ cuts. By making this width--work trade-off explicit and executable,
QWeaver turns circuit cutting from manual preprocessing into a portable
compiler--runtime abstraction and provides a concrete foundation for
integrating improved decompositions, noise-aware partitioning, and future heterogeneous 
multi-QPU resources.

\section*{Acknowledgment}    
This research used resources of the Experimental Computing Laboratory (ExCL) at the Oak Ridge National Laboratory, which is supported by the Office of Science of the U.S. Department of Energy under Contract No. DE-AC05-00OR22725.

This research also used resources of the Oak Ridge Leadership Computing Facility, which is a DOE Office of Science User Facility supported under Contract DE-AC05-00OR22725. In particular, the Quantum Computing User Program (QCUP)~\cite{QCUP} run by OLCF enabled the use of IONQ's backends. The authors thank Michael Sandoval from ORNL and Heidi Nelson-Quillin from IONQ for their technical support. 

\begin{appendices}

\section{Pauli-8 wire cutting and reconstruction overhead}
\label{app:pauli8}

\begin{figure}[t]
  \centering
  \resizebox{0.96\linewidth}{!}{%
  \begin{tikzpicture}[
      font=\small,
      fragment/.style={rectangle, rounded corners=2pt, draw=black!65,
        fill=gray!8, minimum width=2.4cm, minimum height=0.72cm,
        align=center, inner sep=3pt},
      measure/.style={rectangle, rounded corners=2pt, draw=orange!70!black,
        fill=orange!14, minimum width=1.55cm, minimum height=0.72cm,
        align=center, inner sep=3pt},
      prepare/.style={rectangle, rounded corners=2pt, draw=blue!65!black,
        fill=blue!8, minimum width=1.65cm, minimum height=0.72cm,
        align=center, inner sep=3pt},
      term/.style={rectangle, rounded corners=2pt, draw=black!38,
        fill=gray!3, text width=2.55cm, minimum height=1.28cm,
        align=left, inner sep=4pt},
      wire/.style={thick, draw=black!75},
      flow/.style={-{Stealth[length=2mm]}, thick, draw=black!65},
    ]

    \node[fragment] (up) at (1.35,0) {upstream fragment};
    \node[fragment] (down) at (10.65,0) {downstream fragment};
    \draw[wire] (up.east) -- node[above, font=\footnotesize]
      {qubit wire $\mathcal{I}$} (down.west);

    \draw[flow] (6,-0.30) --
      node[right=3pt, fill=white, inner sep=1pt, font=\footnotesize]
      {cut and decompose} (6,-1.02);
    \node[fragment, minimum width=2.05cm] (upcut) at (1.35,-2.00)
      {upstream fragment};
    \node[measure] (ma) at (4.15,-2.00) {measure $M_a$};
    \node[prepare] (sa) at (7.85,-2.00) {prepare $\sigma_a$};
    \node[fragment, minimum width=2.05cm] (downcut) at (10.65,-2.00)
      {downstream fragment};
    \draw[wire] (upcut.east) -- (ma.west);
    \draw[wire, dashed, draw=black!40] (ma.east) --
      node[above=10pt, fill=white, inner sep=1pt, align=center,
        font=\scriptsize]
      {no quantum communication\\signed weight $c_a$} (sa.west);
    \draw[wire] (sa.east) -- (downcut.west);

    \node[font=\footnotesize] at (6,-3.05)
      {Eight choices $(M_a,\,\sigma_a,\,c_a)$ for one cut};
    \node[term] at (1.5,-4.08) {
      \textbf{Identity terms}\\[-1pt]
      $\textcolor{orange!80!black}{I}\to
       \textcolor{blue!70!black}{\lvert0\rangle},\ +\tfrac12$\\
      $\textcolor{orange!80!black}{I}\to
       \textcolor{blue!70!black}{\lvert1\rangle},\ +\tfrac12$};
    \node[term] at (4.5,-4.08) {
      \textbf{$X$ terms}\\[-1pt]
      $\textcolor{orange!80!black}{X}\to
       \textcolor{blue!70!black}{\lvert X_+\rangle},\ +\tfrac12$\\
      $\textcolor{orange!80!black}{X}\to
       \textcolor{blue!70!black}{\lvert X_-\rangle},\ -\tfrac12$};
    \node[term] at (7.5,-4.08) {
      \textbf{$Y$ terms}\\[-1pt]
      $\textcolor{orange!80!black}{Y}\to
       \textcolor{blue!70!black}{\lvert Y_+\rangle},\ +\tfrac12$\\
      $\textcolor{orange!80!black}{Y}\to
       \textcolor{blue!70!black}{\lvert Y_-\rangle},\ -\tfrac12$};
    \node[term] at (10.5,-4.08) {
      \textbf{$Z$ terms}\\[-1pt]
      $\textcolor{orange!80!black}{Z}\to
       \textcolor{blue!70!black}{\lvert Z_+\rangle},\ +\tfrac12$\\
      $\textcolor{orange!80!black}{Z}\to
       \textcolor{blue!70!black}{\lvert Z_-\rangle},\ -\tfrac12$};
  \end{tikzpicture}%
  }
  \caption{Pauli-8 replacement of one cut wire.  The identity channel (top)
    is replaced by an upstream measurement and a downstream state preparation
    that can be executed independently.  The eight cards give the observable,
    prepared state, and signed reconstruction coefficient for every assignment
    $a\in\mathcal{A}$; the sign is applied by the weave task.}
  \label{fig:app-pauli8}
\end{figure}
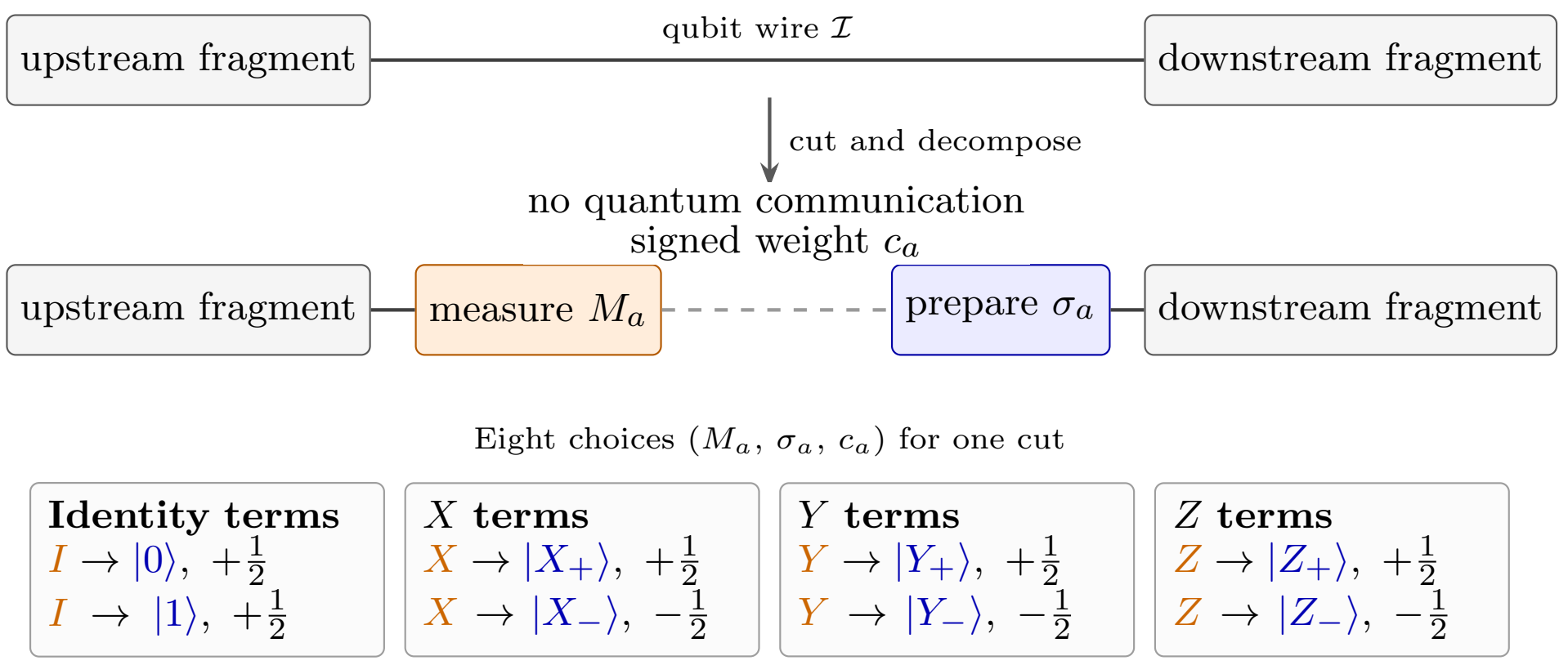

This appendix makes explicit the communication-free wire-cutting identity used
by QWeaver and the resulting experiment and sampling costs.  For a state
$\rho$ at one cut, the single-qubit identity channel is decomposed as
\begin{equation}
  \mathcal{I}(\rho)
  = \frac{1}{2}\operatorname{Tr}(\rho)
      \bigl(\lvert 0\rangle\!\langle 0\rvert
            +\lvert 1\rangle\!\langle 1\rvert\bigr)
    +\frac{1}{2}\sum_{P\in\{X,Y,Z\}}\sum_{s\in\{+1,-1\}}
      s\,\operatorname{Tr}(P\rho)\lvert P_s\rangle\!\langle P_s\rvert,
  \label{eq:app-pauli8-single}
\end{equation}
where $\lvert P_s\rangle$ is the eigenstate of Pauli $P$ with eigenvalue
$s$.  Equation~\eqref{eq:app-pauli8-single} gives the eight signed
measure-and-prepare terms illustrated in Fig.~\ref{fig:app-pauli8}.  The two
$I$ terms require no basis measurement upstream; for the other terms,
$\operatorname{Tr}(P\rho)$ is estimated from the outcomes of a Pauli
measurement.  Negative coefficients are applied only during classical
reconstruction, so no quantum communication is required across the cut
\cite{Peng2020,Mitarai2021}.

For $k$ independently decomposed cuts, an assignment
$\alpha=(a_1,\ldots,a_k)\in\mathcal{A}^{\times k}$ selects one of the eight
terms at every cut and has coefficient
\begin{equation}
  c_\alpha=\prod_{i=1}^{k}c_{a_i},
  \qquad \lvert\mathcal{A}^{\times k}\rvert=8^k.
  \label{eq:app-pauli8-multicut}
\end{equation}
If the selected cuts partition the circuit into $r$ fragments, QWeaver
therefore materializes $8^k$ plan rows and $r8^k$ fragment programs before
pruning or deduplication.  The cut count $k$ and fragment count $r$ are
different graph properties: in a chain-like partition $r=k+1$, but this
relation does not hold for a general circuit, and neither quantity alone fixes
the maximum fragment width.

Let $\widehat{\mu}_{j,\alpha}$ denote the estimate returned by fragment $j$
under assignment $\alpha$.  For a factorizable observable, the reconstructed
estimate is
\begin{equation}
  \widehat{\mu}
    = \sum_{\alpha\in\mathcal{A}^{\times k}} c_\alpha
      \prod_{j=1}^{r}\widehat{\mu}_{j,\alpha};
  \label{eq:app-pauli8-reconstruct}
\end{equation}
output probabilities are reconstructed by the analogous termwise contraction.
This expression is exact in the infinite-shot limit.  To expose the finite-shot
cost, write $\widehat{\mu}_\alpha$ for the complete product in one row and
assume independent row estimates with comparable variance
$\operatorname{Var}(\widehat{\mu}_\alpha)=\sigma^2/S$, where $S$ shots are
acquired per fragment program.  Because all Pauli-8 coefficients have
magnitude $2^{-k}$,
\begin{equation}
  \operatorname{Var}(\widehat{\mu})
    = \sum_\alpha c_\alpha^2
      \operatorname{Var}(\widehat{\mu}_\alpha)
    \simeq \frac{2^k\sigma^2}{S},
  \qquad S_{\mathrm{eff}}=\frac{S}{2^k}.
  \label{eq:app-pauli8-variance}
\end{equation}
Thus, with $S_0$ denoting a cut-independent reference shot count, the total
sampling work scales as
\begin{equation}
  \mathcal{W}_{\mathrm{acq}}(k) =
  \begin{cases}
    O\!\left(rS_0 8^k\right), & \text{fixed shots per fragment program},\\
    O\!\left(rS_0 16^k\right), & \text{fixed statistical precision}.
  \end{cases}
  \label{eq:app-pauli8-work}
\end{equation}
The second line follows because maintaining fixed precision requires
$S=S_0 2^k$.  These are sampling-work bounds for the complete Pauli-8 plan;
backend batching, covariance between estimators, pruning, and more economical
decompositions can change constants or reduce the realized cost.
\end{appendices}

\section*{Declarations}

\begin{itemize}
\item Funding: U.S. Department of Energy, Office of Science under Contract No. DE-AC05-00OR22725. Project: Modular and Error-Aware Software Stack for Heterogeneous Quantum Computing Ecosystems (MACH-Q). 
\item Conflict of interest/Competing interests: Organizational level conflict of interest for being reviewer of this paper. 
\item Ethics approval and consent to participate: Not applicable. This study did not involve human participants or animals.
\item Consent for publication: Not applicable
\item Data availability: The benchmarks used in this study, including the Harris–Ross railway network, are publicly available.
\item Code availability: IRIS/Q-IRIS is publicly available at https://github.com/ORNL/iris. The recent Q-IRIS extensions and the end-to-end QWeaver framework are not yet publicly released; the authors plan to release them as open source after publication.
\item Author contribution: The first four authors contributed equally to different technical aspects of the QWeaver framework and its proof-of-concept implementation. The last two authors led the conceptual discussions and provided research direction.
\item Research Involving Human and /or Animals: Not applicable. This research did not involve humans or animals.
\item Informed Consent: Not applicable
\end{itemize}

\bibliography{papers}

\end{document}